\documentclass[letterpaper, 10pt]{article}

\usepackage{geometry}
\usepackage{graphics} 
\usepackage{epsfig} 
\usepackage{mathptmx} 
\usepackage{times} 
\usepackage{amsthm}
\usepackage{amsmath} 
\usepackage{amssymb, color}  
\usepackage{stmaryrd}
\usepackage{soul}
\newtheorem{theo}{Theorem}
\newtheorem{lem}{Lemma}

\newtheorem{defi}{Definition}
\newtheorem{propo}{Proposition}
\newtheorem{remark}{Remark}
\newtheorem*{pb}{Problem (BHD)}

\usepackage[dvipsnames]{xcolor}

\newcommand{\identity}{\mathbb{I}}
\newcommand{\rref}[1]{(\ref{#1})}
\newcommand{\ical}{\mathcal{I}}

\newcommand{\kk}{\mathcal{K}}
\newcommand{\norme}[1]{\left\Vert #1\right\Vert}
\newcommand{\abs}[1]{\left| #1 \right|}
\newcommand{\reels}{\mathbb{R}}

\newcommand{\entiers}{\mathbb{N}}

\newcommand{\siss}{\text{SISS}}

\newcommand{\pcal}{\mathcal{P}}
\title{\Large\bf
Global stabilization of linear systems with bounds on the feedback and its successive derivatives
}

\author{Jonathan Laporte, Antoine Chaillet and Yacine Chitour 
\thanks{This research was partially supported by a public grant overseen by the French ANR as part of the “Investissements d'Avenir” program, through the iCODE institute, research project funded by the IDEX Paris-Saclay, ANR-11-IDEX-0003-02.}
\thanks{J. Laporte, A. Chaillet  and  Y. Chitour  are with L2S - Univ. Paris Sud - CentraleSup\'elec. 3, rue Joliot-Curie. 91192 - Gif sur Yvette, France.
        {\tt\small jonathan.laporte, antoine.chaillet, yacine.chitour@l2s.centralesupelec.fr}}%
}

\begin{document}
\newcommand{\AC}[1]{\textcolor{blue}{#1}}
\newcommand{\JL}[1]{\textbf{\textcolor{orange}{#1}}}
\newcommand{\ACC}[1]{\textcolor{BlueViolet}{#1}}

\date{}
\maketitle

\section*{Abstract}
We address the global stabilization of linear time-invariant (LTI) systems when the magnitude of the control input and its successive time derivatives, up to an order $p\in\mathbb N$, are bounded by prescribed values. We propose a static state feedback that solves this problem for any admissible LTI systems, namely for stabilizable systems whose internal dynamics has no eigenvalue with positive real part. This generalizes previous work done for single-input chains of integrators and rotating dynamics. 

\section{Introduction}
The study of control systems subject to input constraints is motivated by the fact that signals delivered by physical actuators may be limited in amplitude, and may not evolve arbitrarily fast. An a priori bound on the amplitude of the control signal is usually referred to as \emph{input saturation} whereas a bound on the variation of control signal is referred to as \emph{rate saturation} (e.g \cite{saberi2012}).

Stabilization of linear time-invariant systems (LTI for short) with input saturation has been widely studied in the literature. Such a system is given by $$(S)\ \ \dot x =A x +Bu,$$ where $x\in\mathbb{R}^n$, 
$u$ belongs to a bounded subset of $\mathbb{R}^m$, $A$ is an $n\times n$ matrix and $B$ is an $n\times m$ one. Global stabilization of $(S)$ can be achieved if and only if the LTI system is asymptotically null controllable with bounded controls, i.e., it can be stabilized in the absence of input constraint and the eigenvalues of $A$ have non positive real parts. Saturating a linear feedback law may fail at globally stabilizing $(S)$ as it was observed first in \cite{FULLER69} and then \cite{SY91} for the special case of integrator chains (i.e., when $A$ is the $n$-th Jordan block and $B=(0\cdots 0\ 1)^T$). As shown for instance in \cite{OptCRyan}, optimal control can be used to define a globally stabilizing feedback  for $(S)$ but, when the dimension is greater than $3$, deriving a closed form for this stabilizer becomes extremely difficult. The first globally stabilizing feedback with rather simple closed form (nested saturations) was provided in \cite{Teel92} for chains of integrators and then in \cite{SSY} for the general case. In \cite{Lin95control}, a global feedback stabilizer for $(S)$ was built by relying on control Lyapunov functions arising from a mere existence result. Other globally stabilizing feedback laws for $(S)$ have been proposed with an additional property of robustness with respect to perturbations. In \cite{Saberi:2002ux}, using low-and-high gain techniques, a robust stabilizer was proposed to ensure semiglobal stability, meaning that the control gains can be tuned in such a way that the basin of attraction contains any prescribed compact subset of $\mathbb{R}^n$. This restriction has been removed in \cite{saberi2000}, where the authors provided a global feedback stabilizer for $(S)$ which is robust with respect to perturbations, based on an earlier idea due to Megretsky \cite{Megretski96bibooutput}. Nonetheless, the feedback laws of \cite{saberi2000} and \cite{Megretski96bibooutput} require to solve a nonlinear optimization problem at every point $x\in\mathbb{R}^n$, which makes its practical implementation questionable. In \cite{chitour2015}, an easily implementable global feedback stabilizer for $(S)$ which is robust with respect to perturbations was proposed but it only covers the multiple integrator case and it is discontinuous since it is based on sliding mode techniques. Robust stabilization of $(S)$ was also addressed in \cite{AZCHCHGR15} by relying on the control Lyapunov techniques developed in \cite{Lin95control}.

In contrast to stabilization of LTI systems subject to input saturation, there are much less results available in the literature regarding global stabilization under rate saturation, i.e., when the first time derivative of the control signal is also {\it a priori} bounded. In \cite{Freeman:1998tp}, the authors rely on a backstepping procedure to build a bounded globally stabilizing feedback with a bounded rate, but the methodology does not allow to {\it a priori} impose a prescribed rate. In \cite{SFbound}, a dynamic feedback law inspired from \cite{Megretski96bibooutput} is constructed
and can even be generalized to take into account constraints on higher time derivatives of the control signal. However, as mentioned previously, the numerical efficiency of such feedbacks is definitely questionable.  A rather involved global feedback stabilizer for $(S)$ achieving amplitude and rate saturations was also obtained in \cite{SoSuAL} for continuous time affine systems with a stable free dynamics. This corresponds in our setting to requiring that the matrix $A$ is stable, i.e., $A^T+A\leq 0$ (up to similarity). Finally, let us mention the references \cite{lauvdal97}, \cite{lin1997semi} for semiglobal stabilization results and \cite{SilvaTarbouch03} for local stabilization results using LMIs and anti-windup design. One should also mention \cite{teel1996nonlinear} where a nonlinear small gain theorem is given for the behaviour analysis of control systems with saturation.

The results presented here encompass input and rate saturations as special cases. More precisely, given an integer $p$, we construct a globally stabilizing feedback for $(S)$ such that the control signal and its $p$ first time derivatives, are bounded by arbitrary prescribed positive values, along all trajectories of the closed-loop system. This problem has already been solved by the authors in \cite{LCC1} for the multiple integrator and skew-symmetric cases. The solution given in that paper for the multiple integrator case consisted in considering appropriate nested saturation feedbacks. We also indicated in \cite{LCC1} that these feedbacks fail at ensuring global stability in the skew-symmetric case and we then provided an {\it ad hoc} feedback law for this specific case. Here, we solve the general case with a unified strategy.

The paper should be seen as a first theoretical step towards the global stabilization of an LTI system when the input signal is delivered by a dynamical actuator that limits the control action in terms of magnitude and $p$ first time derivatives. Further developments are needed to explicitly take into account the dynamics of such an actuator. Possible extensions of this work may also address the question of global stabilization by smooth feedback laws (i.e., $C^\infty$ with respect to time) when \emph{all} successive derivatives need to be bounded by prescribed values.

The paper is organized as follows. In Section \ref{sec: main}, we precisely state the problem we want to tackle, the needed definitions as well as the main results we obtain, namely Theorem~\ref{th:main} for the single input case and Theorem~\ref{th:mi} for the multiple input case. Section \ref{sec:proofs} contains the proof of the main results. In section \ref{sec:red:th1} we show that the proof of Theorem \ref{th:main} is a consequence of two propositions. The first one (cf. Proposition~\ref{prop:SISSl}), we show that the feedback proposed in Theorem~\ref{th:main} is indeed a globally stabilizing feedback for $(S)$. We actually prove a stronger result dealing with robustness properties of this feedback, as it is required in \cite{Teel92} and \cite{SSY}. The second proposition (cf. Proposition~\ref{prop:bound:U}) specifically deals with bounding the $p$ first derivatives of the control signal by relying on delicate estimates. Section \ref{sec:red:th2} contains the proof of Theorem \ref{th:mi} which is a consequence of Proposition \ref{prop:SISSl} and Proposition \ref{prop:bound:U:mi}, the latter providing estimates on the successive time derivatives of the control signal. We close the paper by an Appendix, where we gather several technical results used throughout the paper.

\paragraph{Notations :} We use $\reels$ and $\mathbb N$ to denote the sets of real numbers and the set of non negative integers respectively. Given a set $I\subset \mathbb R$ and a constant $a\in\mathbb R$, we let $I_{\geq a}:=\left\{x\in I\,:\, x\geq a\right\}$. Given $m, k \in \entiers$, we define $\llbracket m , k \rrbracket:=\left\{l\in\mathbb N\,:\, l\in[m,k]\right\}$. For a given set $M$,  the boundary of $M$ is denoted by $\partial M$. The factorial of $k$ is denoted by $k!$ and the binomial coefficient is denoted  $\binom{k}{ m}:=\frac{k!}{m! (k-m)!}$.

Given $k\in\mathbb N$ and $n,p\in\mathbb N_{\geq 1}$, we say that a function $f : \reels^n \rightarrow \reels^p$ is of class $C^{k}(\reels^n , \reels^p)$ if its differentials up to order $k$ exist and are continuous, and we use $f^{(k)}$ to denote the $k$-th order differential of $f$. By convention, $f^{(0)}:=f$. 

Given $n,m\in\mathbb N_{\geq 1}$, $\reels^{n,m}$ denotes the set of $n\times m$ matrices with real coefficients. The transpose of a matrix $A$ is denoted by $A^T$. The identity matrix of dimension $n$ is denoted by $\identity_n$. We say that an eigenvalue of $A$ is {\it critical} if it has zero real part and we set $\mu(A):=s(A)+z(A)$ where $s(A)$ is the number of conjugate pairs of nonzero purely imaginary eigenvalues of $A$ (counting multiplicity), and $z(A)$ is the multiplicity of the zero eigenvalue of $A$. We define $A_0 := \begin{pmatrix}
0 & 1 \\
-1 & 0
\end{pmatrix}$, and $b_0 := \begin{pmatrix}
0\\
1 
\end{pmatrix}$. 

We use $\Vert x\Vert$ to denote the Euclidean  norm of an arbitrary vector $x\in\reels^n$. Given $\delta > 0$ and $f : \reels_{\geq 0} \rightarrow \reels^{n}$, we say that $f$ is eventually bounded by $\delta$, and we write $\norme{f(t)} \leq_{ev} \delta$, if there exists $T>0$ such that $\norme{f(t)} \leq \delta$ for all $t \geq T$.
\section{Problem statement and main results}\label{sec: main}

Given $n\in\mathbb N_{\geq 1}$ and $m\in\mathbb N_{\geq 1}$,  consider the LTI system defined by
\begin{equation}
\label{sys:linear}
\dot{x}= Ax +Bu,
\end{equation}
where $x \in \reels^n$, $u \in \reels^m$, $A\in \reels^{n,n}$, and $B\in \reels^{n,m}$. Assume that the pair $(A,B)$ is stabilizable and that all the eigenvalues of $A$ have non positive real parts. Recall that these assumptions on $(A,B)$ are necessary and sufficient for the existence of a bounded continuous state feedback $u=k(x)$ which globally asymptotically stabilizes the origin of (\ref{sys:linear}): see \cite{SSY}. 

Given an integer $p$ and a $(p+1)$-tuple of positive real numbers $(R_j)_{0 \leq j \leq p}$, we want to derive a feedback law whose magnitude and $p$-first time derivatives are bounded by $R_j$, $j\in \llbracket 0 , p \rrbracket$.
\begin{defi}[\emph{feedback law $p$-bounded by $(R_j)_{0 \leq j \leq p}$}]
Given $n\in\mathbb N_{\geq 1}$, $m\in\mathbb N_{\geq 1}$ and $p\in\mathbb N$, let $(R_j)_{0 \leq j \leq p}$ be a $(p+1)$-tuple of positive real numbers. We say that $\nu : \reels^n \rightarrow \reels^m $ is a \emph{feedback law $p$-bounded by $(R_j)_{0 \leq j \leq p}$  for system \rref{sys:linear}} if it is of class $C^p(\reels^n,\reels^m)$ and, for every trajectory of the closed-loop system $\dot{x}= A x +B \nu(x)$, the control signal $U : \reels_{\geq 0} \rightarrow  \reels^m $, $t\mapsto U(t) := \nu (x(t)) $ satisfies $\sup_{t\geq 0} \norme{ U^{(j)}(t)}  \leq R_j $ for all $j \in \llbracket 0 , p \rrbracket$. The function $\nu : \reels^n \rightarrow \reels^m $ is said to be a feedback law $p$-bounded for system \rref{sys:linear}, if there exist $(p+1)$-tuple of positive real numbers $(R_j)_{0 \leq j \leq p}$ such that $\mu(\cdot)$ is a feedback law $p$-bounded by $(R_j)_{0 \leq j \leq p}$  for system \rref{sys:linear}.
\end{defi}

Based on this definition, we can write our stabilization problem of Bounded Higher Derivatives as follows.
\begin{pb}
Given $p\in\mathbb N$ and a $(p+1)$-tuple of positive real numbers $(R_j)_{0 \leq j \leq p}$, design a feedback law $\nu : \reels^n \rightarrow \reels^m$ such that the origin of the closed-loop system $\dot{x}=  A x +B \nu(x)$ is globally asymptotically stable (GAS for short) and the feedback $\nu$ is a feedback law $p$-bounded by $(R_j)_{0 \leq j \leq p}$ for system \rref{sys:linear}.
\end{pb}

Our construction to solve Problem (BHD) will often use the property of \textit{Small Input Small State with linear gain} ($SISS_L$ for short) developed in \cite{SSY}. We recall below its definition


\begin{defi}[\emph{$\siss_L$, \cite{SSY}}]
Given $\Delta>0$ and $N>0$, the control system $\dot{x}=f(x,u)$, with $x\in \reels^n$ and $u\in \reels^m$, is said to be \emph{$\siss_L(\Delta,N)$} if, for all $\delta \in (0, \Delta]$ and all bounded measurable signal $e: \reels_{\geq 0} \rightarrow \reels^m$  eventually bounded by $\delta$, every solution of $\dot{x}=f(x,e)$ is eventually bounded by $N \delta$.
A system is said to be \emph{$\siss_L$} if it is $\siss_L(\Delta , N)$ for some $\Delta,N > 0$. An input-free system $\dot{x}=f(x)$ is called $\siss_L$, if the control system $\dot{x}=f(x)+u$ is $\siss_L$. 
\end{defi}
 
 \begin{remark}\label{rem1}
It follows readily from this definition that if $\dot{x}=f(x)$ is $\siss_L$, then all solutions $\dot{x}=f(x)$ converge to the origin. Note, however, that the $\siss_L$ property does not necessarily ensure GAS in the absence of input, as it does not imply stability of its origin.
\end{remark}

When a feedback law ensures both global asymptotic stability and $\siss_L$, we refer to is an $\siss_L$-stabilizing feedback.

\begin{defi}[\emph{$SISS_L$-stabilizing feedback}]\label{def:sissfeedback}
Given a control system $\dot{x}=f(x,u)$ with $x\in\reels^n$ and $u\in\reels^m$, we say that a feedback law $\nu:\reels^n\to\reels^m$ is \emph{stabilizing} if the origin of the closed-loop system $\dot{x}=f(x,\nu(x))$ is globally asymptotically stable. If, in addition, this closed-loop system is $SISS_L$, then we say that $\nu$ is \emph{ $SISS_L$-stabilizing}.
\end{defi}

As mentioned before the feedback law given in  \cite{LCC1}, which solves Problem (BHD) for the special case of multiple integrators, simply made use of nested saturations with carefully chosen saturation functions. We recall next why this feedback construction cannot work in general. For that purpose it is enough to consider the 2D simple oscillator case which is the control system given by $\dot{x}= \omega A_0 x + b_0 u$, with $x=(x_1, x_2)^T$, $u\in\reels$ and $\omega>0$. This system is one of the two basic systems to be stabilized by means of a bounded feedback, as explained in \cite{SSY}. One must then consider a stabilizing feedback law $u= - \sigma( k^Tx) $, where $k=(k_1,k_2)^T$ is a fixed vector in $\reels^2$ and $\sigma:\reels\rightarrow \reels$ is a saturation function, i.e., a bounded, continuously differentiable function satisfying $s \sigma(s) >0$ for $s \neq 0$ and $\sigma^{(1)}(0)>0$. Note that $k$ is chosen so that the linearized system at $(0,0)$ is Hurwitz. In particular it implies that $k_2\neq 0$. Pick now the following sequence of initial conditions $(l,- k_1l/ k_2)_{l\geq 1}$. A straightforward computation yields that the first time derivative of the control along each trajectory satisfies $\dot{u}(0) = - \sigma^{(1)}(0) \omega l( k_1^2 /k_ 2 + k_2)$, which grows unbounded as $l$ tends to infinity. Therefore this feedback can not be a $1$-bounded feedback.

In order to solve Problem (BHD) for the $2D$ oscillator, we showed in \cite{LCC1} that a feedback law of the type $u_{k,\alpha}:=\frac{k^Tx}{(1+\Vert x\Vert^2)^{\alpha}}$ with $k\in\mathbb{R}^2$ and $\alpha\geq 1/2$ does the job and it also solves Problem (BHD) in case the matrix $A$ in \eqref{sys:linear} is stable. However, we are not able to show whether $u_{k,\alpha}$ stabilizes or not the system in the case where $A := \begin{pmatrix}
A_0 & \identity_2 \\
0 & A_0
\end{pmatrix}$. It turns out that the previous issue is as difficult as asking if a saturated linear feedback stabilizes or not the abovementioned 4D case, which is an open problem.  It is therefore not immediate how to address the general case. This is why Theorem~\ref{th:main} is a non trivial extension of the solution of Problem (BHD) provided for the two-dimensional oscillator.

\subsection{Single input case}

For the case of single input systems the solution of Problem (PHB) is given by the following statement.

\begin{theo}[Single input]
\label{th:main}
Given $n \in \entiers_{>0}$, consider a single input system $\dot{x}=Ax+bu$ where $x \in \reels^n$, $A \in \reels^{n,n}$ and $b\in \reels^{n,1}$. Assume that $A$ has no eigenvalue with positive real part and that the pair $(A,b)$ is stabilizable. Then, given any $p \in \entiers$ and any $(p+1)$-tuple $(R_j)_{0 \leq j \leq p}$ of positive real numbers, there exist 
vectors $k_i\in\reels^n$ and matrices $T_i\in\reels^{n,n}$, $i\in\llbracket1, \mu(A)\rrbracket$, such that the feedback law $\nu:\reels^n\rightarrow\reels$ defined as
\begin{equation}
\label{th:feed}
\nu(x) = - \sum\limits_{j=1}^{\mu(A)} \frac{k_j^T x}{  (1 + \norme{T_l x}^2 )^{1/2}},
\end{equation} 
is a feedback law $p$-bounded by $(R_j)_{0 \leq j \leq p}$ and $SISS_L$-stabilizing for system $\dot{x}=Ax+bu$.
\end{theo} 

In view of Definition \ref{def:sissfeedback}, the feedback law \rref{th:feed} globally asymptotically stabilizes the origin of \rref{sys:linear}, and thus solves Problem (BHD). We stress that, even though the exact computation of the control gains $k_i$ is quite involved (see proof in Section \ref{sec:proofs}), the structure of the proposed feedback law \rref{th:feed} is rather simple. It should also be noted that, unlike the results developed in \cite{LCC1}, this feedback law applies to any admissible single-input systems in a unified manner.

\subsection{Multiple input case}
To give the main result for LTI system with multiple input we need this following definition.
\begin{defi}[Reduced controllability form]
Given $n \in \entiers$ and $q \in \entiers$, a LTI system is said to be in \emph{reduced controllability form} if it reads
\begin{equation}
\label{lin:sys:mi}
\begin{array}{rrr}
\dot{x}_0     =  &  A_{00} x_0  +   A_{01}x_1 +  A_{02}x_2 + \ldots +  A_{0q} x_{q} +  & b_{01} u_1 +  b_{02} u_2 + \ldots +  b_{0q} u_{q} ,  \\
\dot{x}_1     =  &                  A_{11}x_1 +  A_{12}x_2 +  \ldots +  A_{1q} x_{q} + & b_{11} u_1 +  b_{22} u_2 + \ldots +  b_{1q} u_{q}  ,  \\
\dot{x}_2     =  &                               A_{22}x_2 +  \ldots +  A_{2q} x_{q} + &               b_{22} u_2 + \ldots +  b_{2q} u_{q} ,  \\
         \vdots  &                                                                                     &                                                          \\
\dot{x}_{q}  = &                                         A_{qq}x_{q}  + &                          b_{qq}u_{q},  
  \end{array}
\end{equation} where, for some $(q+1)$-tuple $(n_i)_{0\leq i \leq q+1}$ in $\entiers \times (\entiers_{>0})^q$ with $\sum_{i=0}^{q} n_i = n$, $A_{00} \in \reels^{n_0 , n_0}$ is Hurwitz, for every $i \in \llbracket 1, q \rrbracket$ all the eigenvalues of $A_{ii} \in \reels^{n_i , n_i}$ are critical, $b_{ii} \in \reels^{n_i,1}$ and the pairs $(A_{ii}, b_{ii})$ are controllable. 
\end{defi}
From Lemma $5.1$ in \cite{SSY}, it is then clear that without loss of generality, in our case, we can consider that system \rref{sys:linear}  is already given in the reduced controllability form. We can now establish the solution of Problem (BHD) for the multiple input case.

\begin{theo}[Multiple input]
\label{th:mi}
Let $p \in \entiers$ and $(p+1)$-tuple $(R_j)_{0 \leq j \leq p}$ of positive real numbers. Given $n \in \entiers$ and $q \in \entiers$, consider system \rref{lin:sys:mi}. Then, there exist $q$ feedback laws $\kappa_1, \ldots , \kappa_q$ such that:
\begin{itemize}
\item[i)] for every $i \in \llbracket 1, q \rrbracket$, $\kappa_i : \reels^{n_i} \to \reels$ is a feedback law $p$-bounded and $SISS_L$-stabilizing for $\dot{x}_i = A_{ii} x_i +b_{ii} u_i$;
\item[ii)] the feedback law $\mu=[\mu_1, \ldots , \mu_q]^T$ given by 
\begin{eqnarray}
\mu_i  (x_i, \ldots, x_q) & := & \frac{\kappa_i(x_i)}{(1 + \norme{x_{i+1}}^2 + \ldots + \norme{x_q}^2 )^{p+1}} , \quad \forall i \in \llbracket 1, q-1 \rrbracket, \\
\mu_q (x_q) & := & \kappa_q(x_q),
\end{eqnarray} 
is a feedback law $p$-bounded by $(R_j)_{0 \leq j \leq p}$ and $SISS_L$-stabilizing for system \rref{lin:sys:mi}.
\end{itemize}
\end{theo}

This statement provides a unified control law solving Problem (BHD) for all admissible LTI systems. It allows in particular multi-input systems, which was not covered in \cite{LCC1}.

\section{Proof of the main results}\label{sec:proofs}

\subsection{Proof of Theorem \ref{th:main}}
In this section, we prove Theorem \ref{th:main}. For that purpose, we first reduce the argument to establishing of Propositions ~\ref{prop:SISSl} and \ref{prop:bound:U} given below. The first one indicates that the feedback given in Theorem \ref{th:main} is $\siss_L$ stabilizing for $(S)$ in the case of single input. The second proposition provides an estimate of the successive time derivatives of the control signal.
\subsubsection{Reduction of the proof of Theorem \ref{th:main} to the proofs of Propositions~\ref{prop:SISSl} and \ref{prop:bound:U}}\label{sec:red:th1}

Let $n \in \entiers_{\geq 1}$, $p \in \entiers$ and $(R_j)_{0 \leq j \leq p}$ be a $(p+1)$-tuple of positive real numbers. Define $\underline{R}:= \min_{j \in \llbracket 0, p \rrbracket} R_j$. Consider a single input linear system $\dot{x}= Ax +bu$ where $x \in \reels^n$, $A$ and $b$ are $n \times n$ and $n \times 1$ matrices respectively. We assume that the pair $(A,b)$ is stabilizable and that all the eigenvalues of $A$ have non positive real parts. As observed in \cite{SSY}, it is sufficient to consider the case where the pair $(A,b)$ is controllable and all eigenvalues of $A$ are critical. Indeed, since $(A,b)$ is stabilizable there exists a linear change of coordinates transforming $A$ and $b$ into $\begin{pmatrix}A_1\ 0\\0\ \ A_2\end{pmatrix}$ and $ \begin{pmatrix}b_1\\ b_2\end{pmatrix}$,  where $A_1$ is Hurwitz, the eigenvalues of $A_2$ are critical and the pair $(A_2,b_2)$ is controllable. Then, it is immediate to see that we only have to treat the case where $A$ has only critical eigenvalues. From now on, we therefore assume that $A$ has only eigenvalues with zero real parts, and that the pair $(A,b)$ is controllable.

Our construction uses the following linear change of coordinates given by \cite[Lemma 5.2]{SSY}. This decomposition puts the original system in a triangular form made of one-dimensional integrators and two-dimensional oscillators.
\begin{lem}[Lemma $5.2$ in \cite{SSY}]
\label{lem:SSY}
Let $\dot{x}=Ax + b u $, $x \in \reels^n$, $u\in\reels$, be a controllable single input linear system. Assume that all the eigenvalues of $A$ are critical. Let $ \pm i \omega_1, \ldots , \pm i \omega_{s(A)} $ be the nonzero eigenvalues of $A$. Let $(a_2, \ldots , a_{\mu(A)})$ be a family of positive numbers. Define 
\begin{eqnarray}\label{def:theta}
\theta_{i,k}& = & 1  , \quad \text{for} \quad  k=i+1, \nonumber \\ 
\theta_{i,k} & = & \prod_{h=i}^{k-2} 1/a_{h+1}, \quad \text{for} \quad i+2 \leq k \leq \mu(A) +1. 
\end{eqnarray}
Then there is a linear change of coordinates that puts $ \dot{x}=Ax + b u $ in the form
\begin{align}
\dot{y}_i &= \omega_i A_0 y_i + b_0 \sum_{k=i+1}^{s(A)} \theta_{i,k} b_0^T y_k + b_0 \sum_{k=s(A)+1}^{\mu(A)} \theta_{i,k} y_k + \theta_{i,\mu(A)+1}  b_0 u ,\: \quad i=1, \ldots , s(A), \nonumber  \\
\dot{y}_i &= \sum_{k=i+1}^{\mu(A)} \theta_{i,k} y_k + \theta_{i,\mu(A)+1} u, \quad i=s(A)+1 , \ldots , \mu(A)-1,  \label{lem:sys} \\
\dot{y}_{\mu(A)}&= u, \nonumber
\end{align} where $y_i \in \reels^2 $ for $i=1, \ldots , s(A)$ , and $y_i \in \reels$ for $i=s(A)+1 , \ldots , \mu(A)$.
\end{lem}

With no loss of generality, we prove Theorem~\ref{th:main} for system \rref{lem:sys}, where the positive constants $(a_2, \ldots , a_{\mu(A)})$ will be fixed later. Let $a_1 $ be a positive constant. We rely on a candidate feedback $\nu:\reels^n\rightarrow\reels$ under the form 
\begin{equation}
\label{feed:Pj}
\kappa(y) =  - \sum\limits_{i=1}^{s(A)}  \frac{ Q_{i,\mu(A)} b_0^T y_i}{\Big(1 + \sum\limits_{m=i}^{\mu(A)} \norme{y_m}^2 \Big)^{1/2}}  - \sum\limits_{i=s(A)+1}^{\mu(A)}  \frac{Q_{i,\mu(A)} y_i }{\Big( 1 + \sum\limits_{m=i}^{\mu(A)} \norme{y_m}^2 \Big)^{1/2}},
\end{equation} 
with 
\begin{equation}
\label{def:consA}
Q_{i,\mu(A)} := \prod\limits_{l=i}^{\mu(A)} a_l.
\end{equation} 
It therefore remains to choose the positive constants $a_1 , \ldots , a_{\mu(A)}$ such that the feedback law \rref{feed:Pj} is a feedback law $p$-bounded by $(R_j)_{0 \leq j \leq p}$, and  $SISS_L$-stabilizing for system \rref{lem:sys}. For that aim, we rely on the next two propositions, respectively proven in Sections \ref{sec:proofprop1} and \ref{sec:proofprop2}. 

\begin{propo}[]
\label{prop:SISSl}
Let $\dot{x}=Ax + b u $, $x \in \reels^n$, $u\in\reels$, be a controllable single input linear system. Assume that all the eigenvalues of $A$ are critical. Let $ \pm i \omega_1, \ldots , \pm i \omega_{s(A)} $ be the nonzero eigenvalues of $A$. Then, there exist $\mu(A) -1 $ functions $\overline{a}_i : \reels_{>0} \rightarrow \reels_{>0}$, $i \in \llbracket 1 ,\mu(A) -1 \rrbracket$ such that for any constants $a_1 , \ldots , a_{\mu(A)} $ satisfying
\begin{align*}
  a_{\mu(A)} &\in (0, 1], \quad a_i \in (0\,,\, \overline{a}_i (a_{i+1})] , \quad \forall i  \in \llbracket 1 ,\mu(A)-1 \rrbracket,
\end{align*}
 the feedback law \rref{feed:Pj} is $SISS_L$-stabilizing for system \rref{lem:sys}.
\end{propo}

\begin{propo}[]
\label{prop:bound:U}
Let $\dot{x}=Ax + b u $, $x \in \reels^n$, $u\in\reels$, be a controllable single input linear system. Assume that all the eigenvalues of $A$ are critical. Let $ \pm i \omega_1, \ldots , \pm i \omega_{s(A)} $ be the nonzero eigenvalues of $A$. Let $a_i$, $i \in \llbracket 1 , \mu(A) \rrbracket $, be positive constants in $(0,1]$.
Then, there exist a positive constant $c_\mu$, and continuous functions $c_i : \reels^{\mu(A) -i}_{>0 } \to \reels_{>0} $, $i \in \llbracket 1 , \mu(A) -1 \rrbracket$, such that for any trajectory of the closed-loop system \rref{lem:sys} with the feedback law \rref{feed:Pj}, the control signal $U : \reels_{\geq 0} \rightarrow  \reels $ defined by $U(t) := \nu (y(t)) $ for all $t \geq 0$ satisfies, for all $k \in \llbracket 0 ,p \rrbracket $,
\begin{equation*}
\abs{ U^{(k)}(t)} \leq a_\mu c_{\mu(A)}+ \sum\limits_{i=1}^{ \mu(A) - 1 } a_i c_i( a_{\mu(A)} , \ldots , a_{i+1} ), \quad \forall t \geq 0. 
\end{equation*}
\end{propo}

Pick $a_{\mu(A)} \in (0, 1]$ in such a way that
\begin{eqnarray*}
a_{\mu(A)} & \leq & \frac{\underline{R}}{(p+1) c_{\mu(A)}}.
\end{eqnarray*} 
Choose recursively $a_i \in (0, 1]$, $i= \mu(A)- 1 , \ldots , 1$, such that
\begin{eqnarray*}
a_{i} & \leq &  \overline{a}_{i} (a_{i+1}), \quad a_{i}  \leq  \frac{\underline{R}}{(p+1) c_i( a_{\mu(A)} , \ldots , a_{i+1} )},
\end{eqnarray*} 
where the functions $c_i$ appearing above are defined in Proposition~\ref{prop:bound:U}.
By Proposition~\ref{prop:SISSl}, the feedback law \rref{feed:Pj} is $SISS_L$-stabilizing for system \rref{lem:sys}. Moreover, as a consequence of Proposition~\ref{prop:bound:U}, for any trajectory of the closed-loop system \rref{lem:sys} with the feedback law \rref{feed:Pj}, the control signal $U : \reels_{\geq 0} \rightarrow  \reels $ defined by $U(t) := \nu (y(t)) $ for all $t \geq 0$ satisfies $\sup_{t\geq 0}\abs{U^{(k)}(t)} \leq  \underline{R}$ for all $k \in \llbracket 0 ,p \rrbracket $. Thus, the feedback law \rref{feed:Pj} is a  feedback law $p$-bounded by $(R_j)_{0 \leq j \leq p}$ for system \rref{lem:sys}. Since there is a linear change of coordinate ($y=Tx$) that puts \rref{lem:sys} into the original form $\dot{x}=Ax + b u$, the feedback law defined given in \rref{th:feed} can be picked as
\begin{equation*}
\nu(x) : = \kappa(T x)
\end{equation*} 
and it is a feedback law $p$-bounded by $(R_j)_{0 \leq j \leq p}$, and  $SISSL_L$-stabilizing for \eqref{sys:linear}. To sum up, the proof of Theorem~\ref{th:main} boils down to establishing Propositions~\ref{prop:SISSl} and \ref{prop:bound:U}.


\subsubsection{Proof of Proposition \ref{prop:SISSl}}\label{sec:proofprop1}
Proposition \ref{prop:SISSl} is proved by induction on $\mu(A)$. More precisely, we show that the following property holds true for every positive integer $\mu$.
\begin{itemize}
\item[$(P_\mu)$ : ] Given any $\mu \in \entiers_{\geq 1}$, let $s,z \in \entiers $ be such that $s + z = \mu $ and $\omega_1, \ldots , \omega_s$ be positive constants. Then there exist $\mu -1 $ functions $\overline{a}_i : \reels_{>0} \rightarrow \reels_{>0}$, $i \in \llbracket 1 ,\mu -1 \rrbracket$ such that for any constants $a_1 , \ldots , a_\mu $ satisfying
\begin{align*}
  a_{\mu} &\in (0, 1], \quad a_i \in (0\,,\, \overline{a}_i (a_{i+1})] , \quad \forall i  \in \llbracket 1 ,\mu -1 \rrbracket,
\end{align*} 
the feedback law \rref{feed:Pj} is $SISS_L$-stabilizing for system \rref{lem:sys}, with $\mu(A) = \mu$, $s(A) =s$, and $z(A)=z$. Moreover the linearization of this closed-loop system around the origin is asymptotically stable.
\end{itemize}
In order to start the argument, we give intermediate results whose proofs are given in Appendix
and which will be used for the initialization step of the induction and the inductive step. The first statement establishes $SISS_L$ for the one-dimensional integrator.
\begin{lem}
\label{lem:SISS:int}
Let $\epsilon > 1$. For every $\beta >0 $, the scalar system given by
\begin{equation}
\label{sys_int_pert_2}
\dot{x} = - \beta \frac{x} {(1+ x^2)^{1/2}}
\end{equation}
is  $ SISS_L (  \frac{\beta}{2} , \frac{ 2 \epsilon }{ \beta } )$, its origin is $GAS$ and its linearisation around zero is $AS$. 
 \end{lem}
%
The next lemma guarantees that the two-dimensional oscillator is $SISS_L$.
\begin{lem}
\label{Lem:osci}
For every $\omega>0$,  there exist $\Gamma ,N >0$ such that for any $\beta \in (0,1]$ the two-dimensional system given by
\begin{equation}
\label{sys_osc;cor}
\dot{x}= \omega A_0 x - \beta  b_0 \frac{b_0^T x}{(1+ \norme{x}^2)^{1/2}}
\end{equation} 
is $SISS_L(  \beta \Gamma   , \frac{N}{\beta} ) $, its origin is $GAS$ and its linearisation around zero is $AS$. 
\end{lem}
We now start the inductive proof of $(P_{\mu})$. For $\mu=1$, we have to consider two cases. Either $z=1$ and $s=0$ corresponding to the simple integrator
\begin{eqnarray}
\dot{y}_1 = u, \quad \text{with} \quad  u=\kappa(y_1) = - a_1 \frac{y_1}{( 1 + y_1^2)^{1/2}},
\end{eqnarray} or $s=1$ and $z=0$ corresponding to the simple oscillator 
\begin{eqnarray}
\dot{y}_1 = \omega_1 A_0 y_1 + b_0 u,  \quad \text{with} \quad u=\kappa(y_1) = - a_1 \frac{b_0^T y_1}{( 1 + \norme{y_1}^2)^{1/2}},
\end{eqnarray}for some $\omega_1 >0$.
In both cases, $(P_1)$ can be readily deduced by invoking Lemma \ref{lem:SISS:int} and \ref{Lem:osci} respectively. Given $\mu \in \entiers_{>0}$, assume that $(P_\mu)$ holds. In order to establish $(P_{\mu+1})$, it is sufficient to consider the following two cases:
\begin{itemize}
\item[{\bf case i)}] $z = \mu + 1$, i.e, all the eigenvalues of $A$ are zero (multiple integrator);
\item[{\bf case ii)}] $s \geq 1$ , i.e some eigenvalues of $A$ have non zero imaginary part (multiple integrator with rotating modes).
\end{itemize} In both cases we reduce our problem to the choice of only one constant $a_1$ using the inductive hypothesis.
\paragraph{case i)} Let $(a_1 , \ldots , a_{\mu + 1})$ be a set of positive numbers to be chosen later. Consider the multiple integrator given by
\begin{align*}
\dot{y}_i &= \sum_{k=i+1}^{\mu+1 } \theta_{i,k} y_k + \theta_{i,\mu+2} u, \quad i=1 , \ldots , \mu,  \\
\dot{y}_{\mu+1 }&= u, \nonumber
\end{align*} where $y_i \in \reels$ for $i=1 , \ldots , \mu +1 $. Let $\tilde{y}=[ y_2 , \ldots , y_{\mu +1 }]^T$. We then can rewrite this system as 
\begin{align*}
\dot{y}_1 &= \sum_{k=2}^{\mu +1 } \theta_{i,k} y_k + \theta_{i,\mu+2} u,  \\
\dot{\tilde{y}} &=  \tilde{A} \tilde{y} + \tilde{b} u, \nonumber
\end{align*} for some matrices $\tilde{A}$ and $\tilde{b}$ of appropriate dimensions. From the inductive hypothesis, there exist $\mu-1$ functions $\overline{a}_i :\reels_{>0} \rightarrow \reels_{>0}$ for $i \in \llbracket 2 ,\mu  \rrbracket$ such that for any set of positive constants $a_2 , \ldots , a_{\mu +1 } $ satisfying $a_2 , \ldots , a_{\mu +1 } $ satisfying $a_{\mu +1 } \in (0,1]$ and  $0  <  a_i \leq  \overline{a}_i (a_{i +1})$ , for each $ i  \in \llbracket 2 ,\mu \rrbracket$,
 the feedback law $\tilde{\kappa}:\reels^\mu \rightarrow \reels$ defined by 
\begin{equation*}
\tilde{\kappa} (\tilde{y}) =  - \sum\limits_{i=2}^{\mu +1}  \frac{ Q_{i,\mu+1} \: y_i }{(1 + \sum\limits_{m=i}^{\mu+1} \norme{y_m}^2 )^{1/2}}
\end{equation*} is $SISS_L$-stabilizing for $\dot{\tilde{y}} =  \tilde{A} \tilde{y} + \tilde{b} u$. Choose $(a_2 , \ldots , a_{\mu +1 })$ satisfying the above conditions. The feedback law \rref{feed:Pj} is then given by
\begin{equation*}
\kappa(y) = - \tilde{\kappa} (\tilde{y}) - a_1 Q_{2,\mu +1} \frac{y_1}{(1 + \sum\limits_{m=1}^{\mu +1} \norme{y_m}^2 )^{1/2}}  .
\end{equation*} Since $\theta_{1,\mu+2} Q_{k,\mu +1}= \theta_{1,k}$ for all $k \in \llbracket 2, \mu+1 \rrbracket$(see \rref{def:theta} and \rref{def:consA}), the closed-loop system can be rewritten as
\begin{align}
\label{pr:th:SYS:casei}
\dot{y}_1 & =  -  a_1 \frac{y_1}{(1 + \norme{y_1}^2 )^{1/2}} + a_1 \rho_1(y)  +  g_1(\tilde{y})    , \nonumber \\
\dot{\tilde{y}} & = \tilde{A} \tilde{y}  - \tilde{b}  \tilde{\kappa} (\tilde{y})  - \tilde{b} a_1 f_1(y)  , 
\end{align}
 with
\begin{align}
\label{def:ro1:cii}
\rho_1(y) &=  \frac{y_1}{(1 + \norme{y_1}^2 )^{1/2}} \big(1 -  \frac{( 1 + \norme{y_1}^2 )^{1/2}}{ (1 + \sum\limits_{m=1}^{ \mu +1} \norme{y_m}^2 )^{1/2}} \big) , \\ \label{def:g1:cii}
g_1(\tilde{y})& = \sum_{k=2}^{\mu +1} \theta_{1,k} y_k \big( 1 - \frac{1 }{ (1 + \sum\limits_{m=k}^{\mu +1} \norme{y_m}^2 )^{1/2}} \big), \\ \label{def:f1:cii}
f_1(y) & =  \frac{  Q_{2,\mu +1} \: y_1 }{(1 + \sum\limits_{m=1}^{\mu +1} \norme{y_m}^2 )^{1/2}}.
\end{align}
We now move to the other case where the dynamics involves multiple integrators with rotating modes.
\paragraph{case ii)} Let $(a_1 , \ldots , a_{\mu + 1})$ be a set of positive constants to be chosen later. Let $ s \in \entiers_{\geq 1}$, and $ z \in \entiers$ be such that $\mu = s+z$. Let $\omega_1, \ldots , \omega_s$ be a set of non zero real numbers. Consider the following linear control system
\begin{align*}
\dot{y}_i &= \omega_i A_0 y_i + b_0 \sum_{k=i+1}^{s} \theta_{i,k} b_0^T y_k + b_0 \sum_{k=s+1}^{\mu +1 } \theta_{i,k} y_k + \theta_{i,\mu +2 }  b_0 u ,\: \quad i=1, \ldots , s, \nonumber  \\
\dot{y}_i &= \sum_{k=i+1}^{\mu + 1 } \theta_{i,k} y_k + \theta_{i,\mu +2 } u, \quad i=s+1 , \ldots , \mu,  \\
\dot{y}_{\mu +1 }&= u, \nonumber
\end{align*} where $y_i \in \reels^2 $ for $i=1, \ldots , s$ , and $y_i \in \reels$ for $i=s+1 , \ldots , \mu+1$.  Let $\tilde{y}=[ y_2 , \ldots , y_{\mu +1 }]^T$. We then can rewrite this system as follows 
\begin{align*}
\dot{y}_1 &= \omega_1 A_0 y_1 + b_0 \sum_{k=i+1}^{s} \theta_{i,k} b_0^T y_k + b_0 \sum_{k=s+1}^{\mu +1 } \theta_{i,k} y_k + \theta_{i,\mu +2 }  b_0 u ,   \\
\dot{\tilde{y}} &=  \tilde{A} \tilde{y} + \tilde{b} u. \nonumber
\end{align*} From the inductive hypothesis, there exist $\mu-1$ functions $\overline{a}_i :\reels_{>0} \rightarrow \reels_{>0}$ for $i \in \llbracket 2 ,\mu  \rrbracket$  such that for any set of positive constant $a_2 , \ldots , a_{\mu +1 } $ satisfying $a_{\mu +1 } \in (0,1]$ and  $0  <  a_i \leq  \overline{a}_i (a_{i +1})$ , for each $ i  \in \llbracket 2 ,\mu \rrbracket$,
the feedback law $\tilde{\kappa}:\reels^\mu\rightarrow \reels$ defined by
\begin{equation}
\label{feed_int}
\tilde{\kappa}(\tilde{y}) =  - \sum\limits_{i=2}^{s} \frac{Q_{i,\mu+1} \:  b_0^T y_i}{  (1 + \sum\limits_{m=i}^{\mu+1} \norme{y_m}^2 )^{1/2}}  - \sum\limits_{i=s+1}^{\mu+1}   \frac{Q_{i,\mu+1} \:   y_i}{( 1 + \sum\limits_{m=i}^{\mu+1} \norme{y_m}^2 )^{1/2}}
\end{equation} is $SISS_L$-stabilizing for $\dot{\tilde{y}}  = \tilde{A} \tilde{y} + \tilde{b} u$. Choose $a_2 , \ldots , a_{\mu+1}$ satisfying the above conditions. The feedback law \rref{feed:Pj} is then given by
\begin{equation*}
\kappa(y) = - \tilde{\kappa} (\tilde{y}) - a_1 Q_{2,\mu+1} \frac{b_0^T y_1}{(1 + \sum\limits_{m=1}^{\mu +1} \norme{y_m}^2 )^{1/2}}  .
\end{equation*} By noticing that $\theta_{1,\mu+2} Q_{k,\mu+1}= \theta_{1,k}$ for all $k \in \llbracket 2, \mu+1 \rrbracket$  (see \rref{def:theta} and \rref{def:consA}), the closed-loop system can be rewritten as
\begin{align}
\label{pr:th:SYS:caseii}
 \dot{y}_1 & = \omega_1 A_0 y_1  -  a_1 b_0 \frac{b_0^T y_1}{(1 + \norme{y_1}^2 )^{1/2}} + a_1 b_0 \rho_1(y) + b_0 g_1(\tilde{y}) , \nonumber \\
  \dot{\tilde{y}} & = \tilde{A} \tilde{y}  - \tilde{b}  \tilde{\kappa} (\tilde{y})  - \tilde{b}  a_1 f_1(y),
\end{align} with 
\begin{align}
\label{ro_1:est}
\rho_1(y) &= \frac{b_0^T y_1}{(1 + \norme{y_1}^2 )^{1/2}} \big(  1 - \frac{(1 + \norme{y_1}^2 )^{1/2}}{(1 + \sum\limits_{m=1}^{\mu+1} \norme{y_m}^2 )^{1/2}} \big) , \\ \label{g_1:est}
g_1(\tilde{y})& = \sum_{k=2}^{s} \theta_{1,k} b_0^T y_k ( 1 - \frac{1 }{(1 + \sum\limits_{m=k}^{\mu+1} \norme{y_m}^2 )^{1/2}} ) + \sum_{k=s+1}^{\mu+1} \theta_{1,k} y_k ( 1 - \frac{1 }{  (1 + \sum\limits_{m=k}^{\mu+1} \norme{y_m}^2 )^{1/2}} ), \\
\label{f_1:est}
f_1(y) & =   \frac{ Q_{2,\mu+1} \: b_0^T y_1}{ (1 + \sum\limits_{m=1}^{\mu+1} \norme{y_m}^2 )^{1/2}} .
\end{align}

\paragraph{}
In both cases, it remains to show that there exists a function $\overline{a}_1$ such that if $a_1 \in (0, \overline{a_1}]$ then the closed-loop systems \rref{pr:th:SYS:casei} and \rref{pr:th:SYS:caseii}  are $SISS_L$, globally asymptotically stable with respect to the origin, and theirs respective linearizations at zero are asymptotically stable. It is sufficient to prove that the closed-loop systems are $SISS_L$ and their linearization at zero are asymptotically stable. Indeed, from Remark \ref{rem1}, the $SISS_L$ property guarantees the convergence of any solution of the closed-loop with no input. If moreover the linearized system is asymptotically stable, then the globally asymptotic stability of zero follows readily.

For any $a_1>0$, the linearization at zero of the $y_1$-subsystem in \rref{pr:th:SYS:casei} (respectively  \rref{pr:th:SYS:caseii}) is asymptotically stable since it is given by $\dot{y}_1 = - a_1 y_1$ (respectively $\dot{y}_1 = (\omega_1 A_0 - a_1 b_0 b_0^T ) y_1$). Moreover, the linearization at zero of the $\tilde{y}$-subsystem in \rref{pr:th:SYS:casei} (respectively  \rref{pr:th:SYS:caseii}) is given by $\dot{\tilde{y}}= ( \tilde{A}-\tilde{b}\tilde{\kappa}^{(1)}(0)) \tilde{y} - a_1 \tilde{b} y_1  $ (respectively $\dot{\tilde{y}}=  (\tilde{A}-\tilde{b}\tilde{\kappa}^{(1)}(0)) \tilde{y} -  a_1 \tilde{b} b_0^T y_1$). Due to the inductive hypothesis, the origin of $\dot{\tilde{y}}=  \tilde{A}-\tilde{b}\tilde{\kappa}^{(1)}(0)) \tilde{y}$ is asymptotically stable. Thus, local asymptotic stability of \rref{pr:th:SYS:casei} and  \rref{pr:th:SYS:caseii} follows easily.
It remains to prove that systems \rref{pr:th:SYS:casei} and \rref{pr:th:SYS:caseii} are $SISS_L$. In both cases, using that $1-1/(1+s)^{1/2} \leq s$ for all $s \geq 0$, it holds from \eqref{def:g1:cii} and \eqref{g_1:est} that
\begin{equation}
\label{SISS:est:G(y)}
\norme{g_1(\tilde{y})} \leq \sum\limits_{k=2}^{\mu +1} \theta_{1,k} \norme{y_k} \left(   \sum\limits_{m=k}^{\mu +1} \norme{y_m}^2   \right) \leq  \norme{\tilde{y}}^3 \sum\limits_{k=2}^{\mu +1} \theta_{1,k} ,
\end{equation} and from \rref{def:ro1:cii} and \rref{ro_1:est} that
\begin{equation}
\label{SISS:est:ro(y)}
\abs{\rho_1(y)} \leq \norme{\tilde{y}}^2.
\end{equation}

Recall that, due to the inductive hypothesis, $  \dot{\tilde{y}}  = \tilde{A} \tilde{y}  - \tilde{b}  \tilde{k} (\tilde{y})$ is $SISS_L(\tilde{\Delta} , \tilde{N})$ for some $\tilde{\Delta}>0 $ and $\tilde{N}>0$. We next prove the $SISS_L $ property for {\bf case ii)}.

Let 
\begin{eqnarray}
 \label{def:C:cii}
C_1 &: =& \tilde{N} ( Q_{2,\mu+1} \norme{\tilde{b}} + 1 ), \\ \label{def:rho1ov:cii}
C_2 &:=  & C_1^2 + C_1^3 \sum\limits_{k=2}^{\mu +1} \theta_{i,k} .
\end{eqnarray} From Lemma \ref{Lem:osci} (with $\omega = \omega_1$), there exist $\Gamma_1, \: N_1 >0$ such that for any $a_1 \in (0,1]$ the system
$\dot{y}_1 = \omega_1 A_0 y_1  -  a_1 b_0 \frac{b_0^T y_1}{(1 + \norme{y_1}^2 )^{1/2}} $ is $SISS_L( \Gamma_1 a_1  , N_1 / a_1 ) $. Define
\begin{equation}
\label{ch:a1:cii}
\overline{a}_1 := \min \left\lbrace 1 \: ,  \frac{\tilde{\Delta}  \tilde{N}}{C_1 } ,\:  \sqrt{\frac{ \Gamma_1 }{2 C_2}}, \: \sqrt{ \frac{C_1}{4  Q_{2,\mu+1} \tilde{N} \norme{\tilde{b}} N_1 C_2 }}\right\rbrace,
\end{equation}and choose $a_1 \in (0 ,\overline{a}_1 ]$.  Let 
\begin{equation}
\label{ch:delta:cii}
\Delta: =  \min \left\lbrace  \frac{a_1   \Gamma_1 }{2}, a_1 \right\rbrace.
\end{equation}
Given $\delta \leq \Delta$, let $e_1 : \reels_{\geq 0} \rightarrow \reels^2 $ and $e_2 :\reels_{\geq 0}  \rightarrow  \reels^{2s+z-2}$ be two bounded measurable functions, eventually bounded by $\delta$. Consider any trajectory $(y_1(\cdot) , \tilde{y}(\cdot) )$ of the following system
\begin{align}
\label{proof:prop1:SYS:caseii:pert}
 \dot{y}_1 & = \omega_1 A_0 y_1  -  a_1 b_0 \frac{b_0^T y_1}{(1 + \norme{y_1}^2 )^{1/2}} + a_1 b_0 \rho_1(y) + b_0 g_1(\tilde{y}) +e_1, \nonumber \\
  \dot{\tilde{y}} & = \tilde{A} \tilde{y}  - \tilde{b}  \tilde{\kappa} (\tilde{y})  - \tilde{b}  a_1 f_1(y) + e_2,
\end{align}In view of \rref{pr:th:SYS:caseii}, \rref{ro_1:est}, \rref{g_1:est}, \rref{f_1:est} and \rref{feed_int} the above system is clearly forward complete. We next show that there exists a constant $N >0$ such that $\norme{y_1(t)} \leq_{ev} N \delta$ and $\norme{\tilde{y}(t)}\leq_{ev} N \delta$. 
From \rref{f_1:est} and recalling that $\norme{b_0}=1$, a straightforward computation yields
\begin{equation*}
\norme{a_1  \tilde{b} f_1(y)} \leq a_1 Q_{2,\mu +1} \norme{\tilde{b}}.
\end{equation*} Since $\norme{e_2(t) } \leq_{ev} \delta $, it follows that
\begin{equation*}
\norme{a_1  \tilde{b} f_1(y(t)) + e_2(t) }\leq_{ev}  a_1 Q_{2,\mu +1} \norme{\tilde{b}} + \delta.
\end{equation*}Moreover from \rref{ch:a1:cii}, \rref{ch:delta:cii} and  it follows that
\begin{equation*}
\norme{a_1  \tilde{b} f_1(y(t)) + e_2(t) }\leq_{ev}  a_1(Q_{2,\mu +1} \norme{\tilde{b}} + 1) \leq a_1 C_1 / \tilde{N} \leq \tilde{\Delta} ,
\end{equation*} where $C_1$ is defined in \rref{def:C:cii}. Using the $ SISS_L( \tilde{\Delta}, \tilde{N} )$ property of System $\dot{\tilde{y}} = \tilde{A} \tilde{y}  - \tilde{b}  \tilde{\kappa} (\tilde{y})$, it follows  that the solution of \rref{proof:prop1:SYS:caseii:pert} satisfies 
$$\norme{\tilde{y}(t)} \leq_{ev} a_1 C_1. $$ 
Consequently, using \rref{SISS:est:ro(y)} and \rref{SISS:est:G(y)}, it follows that
\begin{equation}
\norme{a_1 b_0 \rho_1(y(t)) + b_0 g_1(\tilde{y}(t))}\leq_{ev} a_1^3 C_2.
\end{equation}
Using \rref{ch:a1:cii}, we have $a_1^3 C_2 \leq  \frac{a_1  \Gamma_1 }{2}$. Moreover \rref{ch:delta:cii} ensures that $\norme{e_1(t)} \leq_{ev }   \frac{a_1  \Gamma_1 }{2}$. So it follows that
\begin{equation*}
\norme{a_1 b_0 \rho_1(y(t)) + b_0 g_1(\tilde{y}(t)) + e_1(t) } \leq_{ev}  a_1  \Gamma_1. 
\end{equation*} The $SISS_L( \Gamma_1 a_1  , N_1 / a_1 ) $ property of $\dot{y}_1 = \omega_1 A_0 y_1  -  a_1 b_0 \frac{b_0^T y_1}{(1 + \norme{y_1}^2 )^{1/2}} $ ensures that
\begin{equation}
\label{est:y_1:cii:1}
\norme{y_1(t) } \leq_{ev} \frac{N_1 }{a_1} (a_1^3 C_2 + \delta)  \leq  N_1 \Gamma_1.
\end{equation}

Now let $\theta >0$ be defined as
\begin{equation}
\label{eq:proof:prop1:limsup}
\theta := \limsup_{t \rightarrow + \infty} \norme{\tilde{y}(t)} .
\end{equation} Then $ \norme{\tilde{y}(t)} \leq_{ev} 2 \theta$. There are two cases to consider, either $2 \theta \leq a_1 C_1$ or $ a_1 C_1 < 2 \theta $. In the case when $2 \theta \leq a_1 C_1$, we have
\begin{equation*}
\norme{a_1 b_0 \rho_1(y(t)) + b_0 g_1(\tilde{y}(t)) + e_1(t) } \leq_{ev}  2 \theta a_1^2 C_2 / C_1.
\end{equation*} So invoking again the $SISS_L(\overline{\rho}_1 \Gamma_1 a_1  , N / a_1 ) $ property of $\dot{y}_1 = \omega_1 A_0 y_1  -  a_1 b_0 \frac{b_0^T y_1}{(1 + \norme{y_1}^2 )^{1/2}} $, one gets that the solution of \rref{proof:prop1:SYS:caseii:pert} satisfies

\begin{equation}
\label{est:y_1:cii:2}
\norme{y_1(t) } \leq_{ev}  \frac{N_1}{a_1} ( \frac{2 \theta a_1^2 C_2}{C_1} + \delta).
\end{equation} In the case when $ a_1 C < 2 \theta $, the estimate \rref{est:y_1:cii:2} follows readily from \rref{est:y_1:cii:1}. Exploiting again the $ SISS_L( \tilde{\Delta}, \tilde{N} )$ property of System $\dot{\tilde{y}} = \tilde{A} \tilde{y}  - \tilde{b}  \tilde{\kappa} (\tilde{y})$, it follows that
\begin{eqnarray*}
\norme{\tilde{y}(t)} & \leq_{ev} & \tilde{N} \left(\norme{\tilde{b}}  Q_{2,\mu +1} N_1 ( \frac{2 \theta a_1^2 C_2}{C_1} + \delta) + \delta \right) \\
& = & \theta \frac{ 2 Q_{2,\mu +1} \tilde{N} \norme{\tilde{b}}  N_1 a_1^2 C_2}{C_1 } + \delta \tilde{N}  ( \norme{\tilde{b}} Q_{2,\mu +1} N_1 + 1 ).
\end{eqnarray*} It then follows from \rref{ch:a1:cii} that
\begin{equation*}
\norme{\tilde{y}(t)}  \leq_{ev} \frac{\theta}{2} +  \delta \tilde{N}  ( \norme{\tilde{b}} Q_{2,\mu +1} N_1 + 1 ).
\end{equation*} Taking the limsup of the above estimate, we get from \rref{eq:proof:prop1:limsup} that
\begin{equation*}
\theta \leq 2 \delta \tilde{N}  ( \norme{\tilde{b}} Q_{2,\mu +1} N_1 + 1 ).
\end{equation*} Consequently, we obtain that
\begin{eqnarray*} 
\norme{\tilde{y}(t)}  & \leq_{ev} & 2  \tilde{N}  ( \norme{\tilde{b}} Q_{2,\mu +1} N_1 + 1 ) \delta ,\\
\norme{y_1(t) } & \leq_{ev} & 2  \frac{N_1}{a_1} ( \frac{2 a_1^2 C_2}{C_1} +1 )  \tilde{N}  ( N_1 + 1 ) \delta,
\end{eqnarray*} which finishes to establish $(P_{\mu +1})$ for the case $ii)$. Proceeding as in case $ii)$, it can be shown that system \rref{pr:th:SYS:casei} is $\siss_L$. This end the inductive proof of $(P_\mu)$. 
\newline


\subsubsection{Proof of Proposition \ref{prop:bound:U}}\label{sec:proofprop2}
Fix $\mu \in \entiers_{\geq 1}$. Let $s$ and $z$ be two integers such that $s + z = \mu $, $\omega_1, \ldots , \omega_s$ be positive constant numbers, and $a_1, \ldots , a_\mu $ be positive numbers less than or equal to 1. Consider the system \rref{lem:sys} with the feedback law \rref{feed:Pj}, where $\mu(A) = \mu$, $s(A) =s$ and $z(A)=z$. We establish Proposition \ref{prop:bound:U} by induction on $p$. More precisely we prove the following statement:
\begin{itemize}
\item[$(H_p)$ :] For each $p \in \entiers$, there exist a positive constant $c_\mu$ and continuous functions $c_i : \reels^{\mu -i}_{>0 } \to \reels_{>0} $, $i \in \llbracket 1 , \mu -1 \rrbracket$, such that for any trajectory $y(\cdot)$ of the closed-loop system \rref{lem:sys} with the feedback law \rref{feed:Pj}, the control signal $U : \reels_{\geq 0} \rightarrow  \reels $ defined by $U(t) := \kappa (y(t)) $ for all $t \geq 0$ satisfies, for all $k \in \llbracket 0 ,p \rrbracket $,
\begin{equation*}
\abs{U^{(k)}(t)} \leq a_\mu c_{\mu}+ \sum\limits_{i=1}^{ \mu - 1 } a_i c_i( a_{\mu} , \ldots , a_{i+1} ), \quad \forall t \geq 0. 
\end{equation*}
\end{itemize}

For $p=0$, this statement ($H_0$) holds trivially. Indeed, it is easy to see that for any trajectory of the closed-loop system \rref{lem:sys} with the feedback law \rref{feed:Pj} we have
\begin{equation*}
\abs{U(t)} \leq a_\mu +  \sum\limits_{i=1}^{\mu-1} a_i Q_{i+1, \mu}, \quad \forall t \geq 0. 
\end{equation*} 
Now, assume that $(H_p)$ holds true for some $p \in \entiers$. We next prove that $(H_{p+1})$ also holds true. To that aim, let $y(\cdot)$ be any trajectory of the closed-loop system \rref{lem:sys} with the feedback law \rref{feed:Pj}, and the control signal $U(t):= \kappa(y(t))$, $\forall t \geq 0 $. By the induction hypothesis, there exist a positive constant $\Upsilon_\mu$ and continuous functions $\Upsilon_i : \reels^{\mu -i}_{>0 } \to \reels_{>0} $, $i \in \llbracket 1 , \mu -1 \rrbracket$, such that for every $k \in \llbracket 0 ,p \rrbracket $  it holds that 
\begin{equation}
\label{fedbound:ind:H}
\abs{U^{(k)}(t)} \leq a_\mu \Upsilon_\mu + \sum\limits_{i=1}^{ \mu - 1 } a_i\Upsilon_i( a_\mu , \ldots , a_{i+1} ),\quad \forall t\geq 0.
\end{equation}
It is sufficient to show that there exist a positive constant $\tilde{\Upsilon}_\mu$ and continuous functions $\tilde{\Upsilon}_i : \reels^{\mu -i}_{>0 } \to \reels_{>0} $, $i \in \llbracket 1 , \mu -1 \rrbracket$, such that
\begin{equation}
\label{fedbound:ind:H:p+1}
\abs{U^{(p+1)}(t)} \leq a_\mu \tilde{\Upsilon}_\mu + \sum\limits_{i=1}^{ \mu - 1 } a_i \tilde{\Upsilon}_i( a_\mu , \ldots , a_{i+1} ),\quad \forall t\geq 0.
\end{equation}
Indeed, the desired results will be obtained by setting $c_\mu := \max \lbrace \Upsilon_\mu , \tilde{\Upsilon}_{\mu} \rbrace$, and $c_i( \cdot) := \max\lbrace \Upsilon_i( \cdot)  , \tilde{\Upsilon}_i( \cdot)   \rbrace$ for $i \in \llbracket 1 , \mu -1 \rrbracket$. In order to establish \eqref{fedbound:ind:H:p+1}, we start by defining the following auxiliary functions: 
\begin{align}\label{g}
g(s):= s^{-1/2}, \quad \forall s > 0
\end{align}
and, for all $t \geq 0$,
\begin{eqnarray}
\label{f_i}
f_i(t) & := &1 +  \sum\limits_{l=i}^{\mu} \norme{y_l(t)}^2, \quad i \in \llbracket 1, \mu \rrbracket . 
\end{eqnarray}
Then, we can rewrite $U(\cdot)$ as 
\begin{equation}
\label{def:U_t}
U(t) = -  \sum\limits_{i=1}^{\mu} U_i(t), \quad \forall t \geq 0,
\end{equation}
where, for every $i \in \llbracket 1 , \mu \rrbracket $,
\begin{eqnarray}
\label{def:U_i_t_1}
U_i(t) & := & Q_{i,\mu} b_{0,i}^T y_i(t) g(f_i(t)), \quad \forall t \geq 0, \label{def:U_i_t_2}
\end{eqnarray} 
where $b_{0,i} = b_0$ for all $i \in \llbracket 1 , s \rrbracket$ and $b_{0,i}=1$ otherwise, and $Q_{i, \mu}$ is defined in \rref{def:consA}. The $(p+1)$-th time derivative of the control signal $U(\cdot)$ is given, for all $t \geq 0$, by $U^{(p+1)}(t)  =  -  \sum_{i=1}^{\mu} U_i^{(p+1)}(t)$. Therefore to prove $(H_{p+1})$, it is sufficient to show that, for each $i \in \llbracket 1 , \mu \rrbracket$, there exists continuous functions $c_{i,l} : \reels^{\mu -l}_{>0 } \to \reels_{>0} $ , $l \in \llbracket 1 , i \rrbracket$, such that, for all $t \geq 0$,
\begin{equation}
\label{bound_U_i_p+1}
\abs{U_i^{(p+1)}(t) } \leq  \sum\limits_{l=1 }^{i} a_l c_{i,l}(a_\mu , \ldots , a_{l+1}),
\end{equation}
$c_{i,\mu}$ is actually a constant independent of $a_\mu$, we write it as $c_{i , \mu} (a_{\mu},a_{\mu+1}) $ for the sake of notation homogeneity.

For $i \in \llbracket 1 , \mu \rrbracket $, we apply Leibniz's rule to \rref{def:U_i_t_1} with respect to $b_{0,i}^T y_i(t) $ and $g(f_i(t))$ and obtain that the $(p+1)$-th time derivative of $U_i(\cdot)$ is given, for all $t \geq 0$, by 
\begin{eqnarray*}
U_i^{(p+1)}(t) & = & a_i Q_{i+1,\mu} \left( \sum\limits_{l_1 = 0}^{p+1 } \binom{p+1}{l_1} b_{0,i}^T y_i^{(p+1-l_1)}(t) [g \circ f_i]^{(l_1)}(t) \right).
\end{eqnarray*} To obtain \rref{bound_U_i_p+1}, it is sufficient to prove that for each $i \in \llbracket 1 , \mu \rrbracket$, and $l_1 \in \llbracket 0, p+1 \rrbracket$ there exist continuous functions $ \beta_{i,l, l_1}  : \reels^{\mu -l}_{>0 } \to \reels_{>0}$ for $l \in \llbracket 1, i \rrbracket $ such that, for all $t \geq 0$,
\begin{eqnarray}
\label{est:int1}
\abs{b_{0,i}^T y_i^{(p+1-l_1)}(t) [g \circ f_i]^{(l_1)}(t)  }  & \leq & \beta_{i,i,l_1}(a_\mu ,\ldots , a_{i+1}) + \sum\limits_{l=1 }^{i-1} a_l \beta_{i,l,l_1}(a_\mu ,\ldots , a_{l+1}).
\end{eqnarray}

In order to get \rref{est:int1} we next provide,  for each $i \in \llbracket 1 , \mu \rrbracket$, estimates of $\|y_i^{(l_1)}(t)\|$, $|f^{(l_1)}_i (t)|$ and $[g \circ f_i]^{(l_1)}(t) $ for $l_1 \in \llbracket 1 , p+1 \rrbracket$. One can observe that, for each $i \in \llbracket 1 , \mu \rrbracket$, $\dot{y}_i$ depends on the constants $a_{i+1}, \ldots , a_\mu$, the states $y_i , \ldots , y_\mu$ and the feedback $u=\kappa(y)$. By an induction argument using differentiation of system \rref{lem:sys}, one can obtain the following statement: for any $k \in \llbracket 1 , p+1 \rrbracket $, $i \in \llbracket 1 , \mu  \rrbracket $, there exist continuous functions 
\begin{eqnarray*}
\Psi_{k,i,l} & : & \reels^{\mu - i}_{>0} \rightarrow \reels_{>0}, \quad l \in \llbracket i+1 , \mu  \rrbracket ,\quad \Phi_{k,i,l} : \reels^{\mu - i}_{>0} \rightarrow \reels_{>0}, \quad l \in \llbracket 0 , p  \rrbracket ,
\end{eqnarray*} such that, for all positive times, it holds that
\begin{equation*}
\norme{y_i^{(k)}(t)} \leq  \sum\limits_{l  =i}^{\mu } \Psi_{k,i,l}(a_\mu , \ldots , a_{i+1}) \norme{y_l(t)} +  \sum\limits_{l  =0}^{k-1} \Phi_{k,i,l}(a_\mu , \ldots , a_{i+1}) \abs{U^{(l)}(t)},
\end{equation*} 
where, by convention, $\Psi_{k,i,\mu}$ are constant functions independent of $a_\mu$ for $k \in \llbracket 1 , p+1 \rrbracket $ and $i \in \llbracket 1 , \mu  \rrbracket $. Using \rref{fedbound:ind:H} in the above estimate, one gets that, for any $k \in \llbracket 1 , p+1 \rrbracket $ and $i \in \llbracket 1 , \mu-1  \rrbracket $, there exist functions $\tilde{v}_{l,k,i} : \reels^{\mu - i}_{>0} \rightarrow \reels_{>0}$, for $ l \in \llbracket i+1 , \mu  \rrbracket $, and $\tilde{\Phi}_{l,k,i} : \reels^{\mu - i}_{>0} \rightarrow \reels_{>0}$ such that, for all $t \geq 0$,
\begin{eqnarray*}
\norme{y_i^{(k)}(t)} & \leq & \sum\limits_{l  =i}^{\mu } \Psi_{k,i,l}(a_\mu , \ldots , a_{i+1}) \norme{y_l(t)} + \tilde{\Phi}_{k,i}(a_\mu , \ldots , a_{i+1})   +  \sum\limits_{l  =1}^{i} a_l  \tilde{v}_{l,k,i}(a_\mu , \ldots , a_{l+1}).
\end{eqnarray*} 
Setting, for $i \in \llbracket 1 , \mu  \rrbracket$, 
\begin{eqnarray*}
\overline{\Psi}_i(a_\mu , \ldots , a_{i+1}) & := & \max \lbrace  \Psi_{k,i,l}(a_\mu , \ldots , a_{i+1}) \: : \: k \in \llbracket 1, p+1 \rrbracket , \:  l \in \llbracket i+1 , \mu \rrbracket \rbrace , \\
\overline{\Phi}_i(a_\mu , \ldots , a_{i+1}) & := & \max \lbrace \tilde{\Phi}_{k,i}(a_\mu , \ldots , a_{i+1} ) \: : \: k \in \llbracket 1, p+1 \rrbracket  \rbrace ,  \\
\tilde{v}_{l ,i} (a_\mu , \ldots , a_{l+1}) & := & \max \lbrace \tilde{v}_{l ,k,i}(a_\mu , \ldots , a_{l+1}) \: : \: k \in \llbracket 1, p+1 \rrbracket  \rbrace , \quad l \in \llbracket 1 , i \rrbracket,
\end{eqnarray*} 
one can obtain that,  for all $k \in \llbracket 1 , p+1 \rrbracket $, all $i \in \llbracket 1 , \mu  \rrbracket $, and all $t\geq 0$,
\begin{equation}
\label{est:y_i_k}
\norme{y_i^{(k)}(t)} \leq  \overline{\Psi}_i(a_\mu , \ldots , a_{i+1})  \sum\limits_{l  =i}^{\mu }  \norme{y_l(t)} + \overline{\Phi}_i(a_\mu , \ldots , a_{i+1}) +  \sum\limits_{l  =1}^{i} a_l  \tilde{v}_{l ,i} (a_\mu , \ldots , a_{l+1}).
\end{equation}
It follows that \rref{est:int1} for $l_1=0$ holds true. For any $i \in \llbracket 1 , \mu  \rrbracket $ and $k \in \llbracket 1 , p+1 \rrbracket $, the $k$-th time derivative of $f_i(\cdot)$, defined in \rref{f_i}, is given, for all $t \geq 0$, by 
\begin{equation*}
f^{(k)}_i (t) = \sum\limits_{l_1 =0}^{k} \binom{k}{l_1} \sum\limits_{l_2 =i}^{\mu} ( y_{l_2}^{(l_1)} (t) )^T  y_{l_2}^{(k-l_1)} (t) .
\end{equation*} Thus, one can get that
\begin{eqnarray*}
\abs{f^{(k)}_i (t) } & \leq & 2 \sum\limits_{l_2 =i}^{\mu} \norme{ y_{l_2} (t) } \norme{  y_{l_2}^{(k)} (t) } + \sum\limits_{l_1 =1}^{k-1} \binom{k}{l_1} \sum\limits_{l_2 =i}^{\mu} \norme{ y_{l_2}^{(l_1)} (t) } \norme{  y_{l_2}^{(k-l_1)} (t) }, \\
& \leq & \sum\limits_{l_2 =i}^{\mu} \left( \norme{ y_{l_2} (t) }^2 + \norme{  y_{l_2}^{(k)} (t) }^2 \right) + \sum\limits_{l_1 =1}^{k-1}  \binom{k}{l_1} \sum\limits_{l_2 =i}^{\mu} \left(  \norme{ y_{l_2}^{(l_1)} (t) }^2 +  \norme{  y_{l_2}^{(k-l_1)} (t) }^2 \right).
\end{eqnarray*}
From \rref{est:y_i_k}, and using the fact that $\left( \sum\limits_{i_1=1}^{m} \abs{x_{i_1} } \right)^2 \leq m \sum\limits_{i_1=1}^m  x_{i_1}^2 $, one can obtain that for each $l_2 \in \llbracket 1 , \mu \rrbracket $ and $l_1 \in \llbracket 1 , p+1 \rrbracket $ it holds that, for all $t \geq 0 $,
\begin{equation}
\label{est:y_i_k_2}
\norme{ y_{l_2}^{(l_1)} (t) }^2 \leq (\mu +2) \left( \overline{\Psi}_{l_2} (a_\mu , \ldots , a_{l_2 +1})^2   \sum\limits_{l=l_2 }^{\mu} \norme{ y_{l} (t) }^2 +   \overline{\Phi}_{l_2}(a_\mu , \ldots , a_{l_2 +1})^2 +  \sum\limits_{l  =1}^{l_2} ( a_l  \tilde{v}_{l ,l_2} (a_\mu , \ldots , a_{l+1}) )^2\right).
\end{equation} 
Since the right-hand side of \rref{est:y_i_k_2} is independent of $l_1$, and $a_l \leq 1$ for all $l \in \llbracket 1, \mu \rrbracket $, one can gets that there exist continuous functions 
\begin{eqnarray*}
\tilde{\overline{\Psi}}_{l} & : & \reels^{\mu - l}_{>0} \rightarrow \reels_{>0}, \quad l \in \llbracket 1 , \mu  \rrbracket ,\\
 \tilde{\overline{\Phi}}_{l} & : & \reels^{\mu - l}_{>0} \rightarrow \reels_{>0}, \quad l \in \llbracket 1 , \mu  \rrbracket ,\\
 \tilde{\overline{v}}_{l,l_1}  &:& \reels^{\mu - l}_{>0} \rightarrow \reels_{>0}, \quad l_1 \in \llbracket 1 , \mu  \rrbracket, \: l \in \llbracket 1 , l_1  \rrbracket ,
\end{eqnarray*} such that, for any $k \in \llbracket 1, p  \rrbracket $ and all $t \geq 0$, it holds
\begin{equation*}
\abs{f^{(k)}_i (t) } \leq  \tilde{\overline{\Psi}}_{l_2} (a_\mu , \ldots , a_{i +1})  \sum\limits_{l  =i}^{\mu }  \norme{y_l(t)}^2  +  \tilde{\overline{\Phi}}_{l_2}(a_\mu , \ldots , a_{i +1}) +   \sum\limits_{l  =1}^{i}  a_l \tilde{\overline{v}}_{l,i} (a_\mu , \ldots , a_{l+1}).
\end{equation*} 
A trivial estimate for any $k \in \llbracket 1 , p+1 \rrbracket $, any $i \in \llbracket 1 , \mu  \rrbracket $, and all $t\geq 0$ is given by
\begin{equation}
\label{est:f_i_k}
\abs{f^{(k)}_i (t) } \leq   \tilde{\overline{\Psi}}_{i} (a_\mu , \ldots , a_{i +1})  f_{i}(t)  +  \tilde{\overline{\Phi}}_{i}(a_\mu , \ldots , a_{i +1}) +   \sum\limits_{l  =1}^{i}  a_l \tilde{\overline{v}}_{l,l_2} (a_\mu , \ldots , a_{l+1}) .
\end{equation}


By the Fa\`a di Bruno's formula (given in Lemma \ref{lem:fa_di} in Appendix), for each $i \in \llbracket 1 , \mu \rrbracket$, and $l_1 \in \llbracket 1, p+1 \rrbracket$, the $l_1$-th time derivative of $g \circ f_i( \cdot)$ is given, for all $t \geq 0$, by
\begin{eqnarray*}
[g \circ f_i]^{(l_1)}(t) & = & \sum\limits_{l_2=1}^{l_1 } g^{(l_2)}(f_i(t) ) \sum\limits_{\delta \in \pcal_{l_1 ,l_2}} c_\delta \prod\limits_{l=1}^{l_1-l_2+1}(f_i^{(l)}(t))^{\delta_l}, 
\end{eqnarray*} 
where $\pcal_{l_1,l_2}$ denotes the set of $(l_1 - l_2 +1)-$tuples $\delta :=(\delta_1 , \delta_2, \ldots , \delta_{l_1 - l_2 +1})$  of positive integers satisfying $\delta_1 + \delta_2 + \ldots +\delta_{l_1 - l_2+1} = l_2$ and $\delta_1 +2 \delta_2 + \ldots + (l_1 - l_2+1) \delta_{l_1 - l_2+1} =l_1$.
Observe that the $k$-th derivative of the function $g$ defined in \eqref{g} reads
\begin{equation}
\label{eq:dev_g}
g^{(k)}(s) = d_k s^{-1/2 - k}, \quad \forall s>0,
\end{equation}
with $d_k = (-1)^k\prod\limits_{l=0}^{k-1} (1/2 + l )$. Using \rref{eq:dev_g}, and taking the absolute value, one can get, for all $t \geq 0$,
\begin{eqnarray*}
\abs{[g \circ f_i]^{(l_1)}(t)} & \leq & \sum\limits_{l_2=1}^{l_1 } d_{l_2} \frac{1 }{( f_i(t) )^{l_2 + 1/2 }} \sum\limits_{\delta \in \pcal_{l_1 ,l_2}} c_\delta \prod\limits_{l=1}^{l_1-l_2+1}\abs{f_i^{(l)}(t)}^{\delta_l} .
\end{eqnarray*}
Using \rref{est:f_i_k}, one can obtain that, for any $l_1 \in \llbracket 1 , p+1 \rrbracket$, any $l_2 \in \llbracket 1 , l_1 \rrbracket $ and for all $t \geq 0$, 
\begin{eqnarray*}
\sum\limits_{\delta \in \pcal_{l_1 ,l_2}} c_\delta \prod\limits_{l=1}^{l_1-l_2+1}\abs{f_i^{(l)}(t)}^{\delta_l} & \leq & \bigg(  \tilde{\overline{\Psi}}_{i}(a_\mu , \ldots , a_{i +1}) f_{i}(t)+ \tilde{\overline{\Phi}}_{i}(a_\mu , \ldots , a_{i +1})  \\& + &   \sum\limits_{l_3  =1}^{i}  a_{l_3} \tilde{\overline{v}}_{l_3,i} (a_\mu , \ldots , a_{i +1})    \bigg)^{l_2} \sum\limits_{\delta \in \pcal_{l_1 ,l_2}} c_\delta  .
\end{eqnarray*}
It follows that, for all $l_1 \in \llbracket 1 , p+1 \rrbracket$,  $t \geq 0$,
\begin{eqnarray*}
\abs{[g \circ f_i]^{(l_1)}(t)} & \leq & \sum\limits_{l_2=1}^{l_1 } d_{l_2} \frac{\sum\limits_{\delta \in \pcal_{l_1 ,l_2}} c_\delta   }{( f_i(t) )^{1/2 }} \Bigg(  \\
 & &  
\frac{ \tilde{\overline{\Psi}}_{i}(a_\mu , \ldots , a_{i +1}) f_{i}(t)+ \tilde{\overline{\Phi}}_{i}(a_\mu , \ldots , a_{i +1})  +  \sum\limits_{l_3  =1}^{i}  a_{l_3} \tilde{\overline{v}}_{l_3,i} (a_\mu , \ldots , a_{i +1})  }{f_{i}(t) }  \Bigg)^{l_2}, \\
& \leq & \sum\limits_{l_2=1}^{l_1 } d_{l_2} \frac{\sum\limits_{\delta \in \pcal_{l_1 ,l_2}} c_\delta   }{( f_i(t) )^{1/2 }} \Bigg(  \\
 & &  
 \tilde{\overline{\Psi}}_{i}(a_\mu , \ldots , a_{i +1}) + \tilde{\overline{\Phi}}_{i}(a_\mu , \ldots , a_{i +1})  +  \sum\limits_{l_3  =1}^{i}  a_{l_3} \tilde{\overline{v}}_{l_3,i} (a_\mu , \ldots , a_{i +1}) \Bigg)^{l_2}, 
\end{eqnarray*}

Thus, it can be seen that, for every $i \in \llbracket 1 , \mu \rrbracket$ and $l_1 \in \llbracket 1, p+1 \rrbracket$, there exist continuous functions $\Gamma_{i,l_1} : \reels_{>0}^{\mu -i} \rightarrow \reels_{>0} $ and $\Gamma_{i,l_1,l} :  \reels_{>0}^{\mu -l} \rightarrow \reels_{>0} $, $l \in \llbracket 1 , i+1  \rrbracket$, such that, for all $t \geq 0$,
\begin{equation}
\label{est:gfi}
\abs{[g \circ f_i]^{(l_1)}(t)}  \leq \frac{1 }{\big( f_i(t) \big)^{ 1/2 }} \left( \Gamma_{i,l_1}(a_\mu , \ldots , a_{i+1} ) + \sum\limits_{l=1}^{i} a_i \Gamma_{i,l_1,l}(a_\mu , \ldots, a_{i+1})\right).
\end{equation}
Then, from \rref{est:gfi} and \rref{est:y_i_k} it follows that \rref{est:int1} holds true for any $l_1 \in \llbracket 1, p+1 \rrbracket$. This ends the inductive proof of $(H_p)$.
\subsection{Proof of Theorem \ref{th:mi}}
\subsubsection{Reduction of the proof of Theorem \ref{th:mi} to the proof of Propositions \ref{prop:SISSl} and \ref{prop:bound:U:mi}}\label{sec:red:th2}
We prove Theorem \ref{th:mi} by induction on the number of inputs $q$. We show that the inductive step reduces to Proposition \ref{prop:SISSl} and Proposition \ref{prop:bound:U:mi} which is proven in Section \ref{sss:Pr:3}.

For $q=1$, the conclusion follows from Theorem \ref{th:main}. For a given $q \in \entiers_{\geq 1}$ assume that Theorem \ref{th:mi} holds. We show that Theorem \ref{th:mi} then holds for LTI systems given in the reduced controllability form with $q+1$ inputs. Let $p \in \entiers$ and $(R_j)_{0 \leq j \leq p}$ be a $(p+1)$-tuple of positive real numbers. Define $\underline{R}:= \min_{j \in \llbracket 0, p \rrbracket} R_j$. Given $n \in \entiers_{\geq 2}$ consider a LTI system given in the reduced controllability form with $\tilde{q}:=q+1$ inputs by
\begin{equation*}
\begin{array}{rrr}
\dot{x}_0     =  &  A_{00} x_0  +   A_{01}x_1 +  A_{02}x_2 + \ldots +  A_{0\tilde{q}} x_{\tilde{q}} +  & b_{01} u_1 +  b_{02} u_2 + \ldots +  b_{0\tilde{q}} u_{\tilde{q}} ,  \\
\dot{x}_1     =  &                  A_{11}x_1 +  A_{12}x_2 +  \ldots +  A_{1\tilde{q}} x_{\tilde{q}} + & b_{11} u_1 +  b_{22} u_2 + \ldots +  b_{1\tilde{q}} u_{\tilde{q}}  ,  \\
\dot{x}_2     =  &                               A_{22}x_2 +  \ldots +  A_{2\tilde{q}} x_{\tilde{q}} + &               b_{22} u_2 + \ldots +  b_{2\tilde{q}} u_{\tilde{q}} ,  \\
         \vdots  &                                                                                     &                                                          \\
\dot{x}_{\tilde{q}}  = &                                         A_{\tilde{q}\tilde{q}}x_{\tilde{q}}  + &                          b_{\tilde{q} \tilde{q}}u_{\tilde{q}},  
  \end{array}
\end{equation*}
where $x_i \in \reels^{n_i}$ and $u_i \in \reels$ for each $i \in \llbracket 0, q+1 \rrbracket$, $A_{00}$ is Hurwitz, for every $i \in \llbracket 1, q+1 \rrbracket$ all the eigenvalues of $A_{ii} $ are critical, and the pairs $(A_{ii}, b_{ii})$ are controllable. 

Since $A_{00}$ is Hurwitz, if we find a  feedback law $p$-bounded by $(R_j)_{0 \leq j \leq p}$, and $SISS_L$-stabilizing for $(x_1, \ldots , x_{q+1})-$subsystem then, clearly, this feedback does the job for the complete system. From now on, we only consider the $(x_1, \ldots , x_{q+1})-$subsystem and we rewrite it compactly as 
\begin{subequations}
\label{sys:mi}
\begin{align}
\dot{x}_1 &  = A_{11} x_1 + b_{11} u_1 + \tilde{A}z+ \tilde{B}\overline{u}, \label{sys:mi:x} \\
\dot{z} & =  \overline{A} z + \overline{B} \overline{u}, \label{sys:mi:z}
\end{align}
\end{subequations} where $z:=[x_2, \ldots , x_{q+1}]^T$, $u:=[u_2 , \ldots , u_{q+1}]^T$.

We next provide a key technical lemma.
\begin{lem}
\label{lem:struct:l}
Let $\dot{x}=Ax + b u $, $x \in \reels^n$, $u\in\reels$, be a controllable single input linear system. Assume that all the eigenvalues of $A$ are critical. Let $ \pm i \omega_1, \ldots , \pm i \omega_{s(A)} $ be the nonzero eigenvalues of $A$, $(a_2, \ldots , a_{\mu(A)})$ be a sequence of positive numbers and $T \in \reels^{n,n}$ be such that the linear change of coordinate $y=Tx$ transforms $\dot{x}  = A x + bu$ into system \rref{lem:sys} compactly written as $\dot{y}=J y + b u$. Rewrite $T$ as $$T=[T_1 , \ldots , T_{s(A)}, T_{s(A)+1}, \ldots, T_{\mu(A)}]^T,$$ where $T_i \in \reels^{2,n}$ if $i \in \llbracket 1, s(A) \rrbracket$ otherwise $T_i \in \reels^{1,n}$. 
Then $T$ has the following property
\begin{itemize}
\item[$(\ical)$ :]  $T_{\mu(A)}$ is independent of $(a_2, \ldots , a_{\mu(A)})$, and each $T_i$ depend only on $(a_{i+1}, \ldots , a_{\mu(A)})$.
\end{itemize}
 Moreover, given $r,k \in \entiers$, let $M \in \reels^{n,r}$ be independent of the constants $a_i$, then the matrices $T M$ and $J^k T$ satisfy property $(\ical)$. 
\end{lem} The proof of Lemma \ref{lem:struct:l} follows from a careful examination of the proofs of Lemmas $3.1$ and $5.1$ in \cite{SSY}.

\paragraph{}

Let $(a_2, \ldots , a_{\mu(A_{11})})$ be a sequence of positive numbers (to be chosen later). Let $T$ be the linear change of coordinate that transforms $\dot{x}  = A_{11} x + b_{11} u_1$ into the form of system \rref{lem:sys} compactly written as $\dot{y}=J y + b u$. We now make the following changes of coordinates $y= Tx$, and system \rref{sys:mi} is then given by
\begin{subequations}
\label{sys:mi:g}
\begin{align}
\dot{y}&  = J y +b  u_1 + T \tilde{A}z+ T \tilde{B}\overline{u}, \label{sys:mi:x:g} \\
\dot{z} & =  \overline{A} z + \overline{B} \overline{u}. \label{sys:mi:z:g}
\end{align}
\end{subequations}
Let $\kappa$ be a feedback law $p$-bounded feedback law by $(R_j/2)_{0 \leq j \leq p}$, and $SISS_L(N_2, \Delta_2)$-stabilizing for subsystem \rref{sys:mi:z:g}, for some $N_2, \Delta_2 >0$ (thanks to the inductive hypothesis, we know that this feedback exists). Let $a_1>0$, to be chosen later.  We seek the following feedback:
\begin{subequations}
\label{feed:mi:g}
\begin{align}
\label{feed:mi_1}
u_1(y,z)& :=\frac{\mu(y)}{(1+\norme{z}^2)^{p}}, \\ \label{feed:mi_2}
\overline{u}(z) & := \kappa(z), 
\end{align}
\end{subequations} where $\mu(y)$ is defined in \rref{feed:Pj}. We now show that there exist positive constants $(a_1,a_2, \ldots , a_{\mu(A_{11})})$  such that the feedback law \rref{feed:mi:g} is a feedback law $p$-bounded and $SISS_L$-stabilizing for system \rref{sys:mi:g}. This choice is based on Proposition \ref{prop:SISSl} and the following statement which is proven in Section \ref{sss:Pr:3}.
\begin{propo}[\emph{$p$-bounded feedback}]
\label{prop:bound:U:mi}
Let $a_i$, for $i \in \llbracket 1 ,\mu(A_{11}) \rrbracket $, be positive constants in $(0,1]$. Consider system \rref{sys:mi:g} with the feedback law \rref{feed:mi:g}. Assume that $\kappa$ is a feedback law $p$-bounded by $(R_j/2)_{0 \leq j \leq p}$, and $SISS_L(N_2, \Delta_2)$-stabilizing for subsystem \rref{sys:mi:z:g}. 
Then, there exist a positive constant $c_{\mu(A_{11})}$, and continuous functions $c_i : \reels^{\mu(A_{11}) -i}_{>0 } \to \reels_{>0} $, $i \in \llbracket 1 , \mu(A_{11}) -1 \rrbracket$, such that for any trajectory of the closed-loop system \rref{sys:mi:g} with the feedback law \rref{feed:mi:g}, the control signal $U_1 : \reels_{\geq 0} \rightarrow  \reels $ defined by $U_1(t) := u_1 (y(t),z(t)) $ for all $t \geq 0$ satisfies, for all $k \in \llbracket 0 ,p \rrbracket $,
\begin{equation*}
\abs{ U_1^{(k)}(t)} \leq a_\mu c_{\mu(A_{11})}+ \sum\limits_{i=1}^{ \mu(A_{11}) - 1 } a_i c_i( a_{\mu(A_{11})} , \ldots , a_{i+1} ), \quad \forall t \geq 0. 
\end{equation*}
\end{propo}
Pick $a_{\mu(A_{11})} \in (0, 1]$ in such a way that
$$
a_{\mu(A_{11})} \leq \frac{ \underline{R}}{2 (p+1) c_{\mu(A_{11})}}.
$$
Choose recursively $a_i \in (0, 1]$, $i= \mu(A_{11})- 1 , \ldots , 1$, such that
\begin{eqnarray*}
a_{i} & \leq &  \overline{a}_{i} (a_{i+1}), \quad a_{i}  \leq  \frac{\underline{R}}{2(p+1) c_i( a_{\mu(A)} , \ldots , a_{i+1} )},
\end{eqnarray*} 
where the functions $c_i$ appearing above are defined in Proposition~\ref{prop:bound:U:mi} and the functions $ \overline{a}_{i}$ are defined in Proposition \ref{prop:SISSl}.
By Proposition~\ref{prop:SISSl}, the feedback law $\mu(y)$ is $SISS_L$-stabilizing for system $\dot{x}=Jx+bu$. We now prove that the closed-loop system \rref{sys:mi:g} with the feedback \rref{feed:mi:g} is $SISS_L$ (now, all the coefficients have been chosen). To that aim, first notice that there exist $\alpha_1, \alpha_2 >0$ such that, for all $\norme{z}\leq 1$,
\begin{eqnarray*}
\norme{ T \tilde{A}z+ T \tilde{B}\kappa(z) } & \leq & \alpha_1 \norme{z},\\
\norme{b \mu(y) \left( 1 - \frac{1}{(1+\norme{z}^2)^p}\right)} &\leq & \alpha_2 \norme{z}.
\end{eqnarray*}  
Let $$\Delta := \min \left\lbrace 1,\: \Delta_2, \: \frac{1}{N_2} ,\: \frac{\Delta_1}{(\alpha_2 + \alpha_1) N_2 +1 }  \right\rbrace.$$
Given $\delta \leq \Delta$, let $e_1, e_2 $ be two bounded measurable functions of the appropriate dimension, eventually bounded by $\delta$. Consider any trajectory $(y(\cdot) , z(\cdot) )$ of the following system
\begin{align}
\dot{y}&  = J y +b  \mu(y)- b \mu(y) \big( 1 - \frac{1}{(1+\norme{z}^2)^p}\big)  + T \tilde{A}z+ T \tilde{B}\kappa(z)+e_1, \\
\dot{z} & =  \overline{A} z + \overline{B} \kappa(z)+e_2, 
\end{align} From the $SISS_L(\Delta_2, N_2)$ property of $z$-subsystem it follows that $\norme{z(t)} \leq_{ev} N_2 \delta \leq 1$. Thus, using the above estimate, it is immediate to see that $$\norme{b \mu(y(t)) \big( 1 - \frac{1}{(1+\norme{z(t)}^2)^p}\big)  + T \tilde{A}z(t)+ T \tilde{B}\kappa(z(t))+e_1(t)} \leq_{ev} \delta \big((\alpha_1 + \alpha_2) N_2 +1 \big) \leq \Delta_1. $$ Therefore, invoking the $SISS_L(\Delta_1, N_1)$ property of
$\dot{x}=Jx+b\mu(y)$, it follows that $\norme{y(t)}\leq_{ev}\delta \big((\alpha_1 + \alpha_2) N_2 +1 \big) N_1$. So, the closed-loop system \rref{sys:mi:g} with the feedback \rref{feed:mi:g} is $SISS_L$.
Moreover, as a consequence of Proposition~\ref{prop:bound:U:mi} and of the inductive hypothesis, for any trajectory of the closed-loop system \rref{lem:sys} with the feedback law \rref{feed:mi:g}, the control signal $U : \reels_{\geq 0} \rightarrow  \reels^m $, defined by $U(\cdot) := [ U_1(\cdot), U_2(\cdot) ]^T$ with $U_1(t) := u_1 (y(t),z(t)) $ and $U_2(t) := \kappa(z(t))$ for all $t \geq 0$, satisfies $$\sup_{t\geq 0}\norme{U^{(k)}(t)} \leq R_k$$ for all $k \in \llbracket 0 ,p \rrbracket $. Thus, the feedback law \rref{feed:mi:g} is a  feedback law $p$-bounded by $(R_j)_{0 \leq j \leq p}$ for system \rref{sys:mi:g}.

\subsubsection{Proof of Proposition \ref{prop:bound:U:mi}}
\label{sss:Pr:3}
For the sake of notation compactness let $\mu = \mu(A_{11})$. To prove Proposition \ref{prop:bound:U:mi}, we establish by induction on $k$ that the following property holds, for all $k \in \llbracket 0,p \rrbracket$:
\begin{itemize}
\item[$(\overline{H}_k)$ :]There exist a positive constant $c_\mu$, and continuous functions $c_i : \reels^{\mu -i}_{>0 } \to \reels_{>0} $, $i \in \llbracket 1 , \mu -1 \rrbracket$, such that for any trajectory of the closed-loop system \rref{sys:mi:g} with the feedback law \rref{feed:mi:g}, the control signal $U_1 : \reels_{\geq 0} \rightarrow  \reels $ defined by $U_1(t) := u_1 (y(t),z(t)) $ for all $t \geq 0$ satisfies, for all $j \in \llbracket 0 ,k \rrbracket $,
\begin{equation*}
\abs{ U_1^{(j)}(t)} \leq a_\mu c_{\mu}+ \sum\limits_{i=1}^{ \mu - 1 } a_i c_i( a_{\mu} , \ldots , a_{i+1} ), \quad \forall t \geq 0. 
\end{equation*}
\end{itemize}

For $k=0$, the statement ($\overline{H}_0$) holds trivially. Now, assume that $(\overline{H}_k)$ holds true for some $k \in \llbracket 0, p-1 \rrbracket$. We next prove that $(\overline{H}_{k+1})$ also holds true. Let $(y(\cdot),z(\cdot))$ be any trajectory of the closed-loop system \rref{sys:mi:g} with the feedback law \rref{feed:mi:g}, and the control signal $U_1(t):= u_1 (y(t),z(t)))$ and $U_2(t):= \kappa(z(t))$, $\forall t \geq 0 $. As in the proof of Proposition \ref{prop:bound:U}, it is sufficient to prove that there exist a positive constant $\tilde{\Upsilon}_\mu$ and continuous functions $\tilde{\Upsilon}_i : \reels^{\mu -i}_{>0 } \to \reels_{>0} $, $i \in \llbracket 1 , \mu -1 \rrbracket$, such that
\begin{equation}
\label{fedbound:ind:H:k+1}
\abs{U_1^{(k+1)}(t)} \leq a_\mu \tilde{\Upsilon}_\mu + \sum\limits_{i=1}^{ \mu - 1 } a_i \tilde{\Upsilon}_i( a_\mu , \ldots , a_{i+1} ),\quad \forall t\geq 0.
\end{equation}Let $\tilde{q}(s):= s^{-(p+1)}$, for all $s>0$. Define $h(t):= 1 + \norme{z(t)}^2$, for all $t \geq 0$. With the same notation given in the proof of Proposition \ref{prop:bound:U}, one can write $U_1(\cdot)$ as
\begin{equation}
\label{def:U_t:mi}
U_1(t) = -  \sum\limits_{i=1}^{\mu} U_{1i}(t), \quad \forall t \geq 0,
\end{equation}
where, for every $i \in \llbracket 1 , \mu \rrbracket $,
\begin{eqnarray}
\label{def:U_i_t_1:mi}
U_{1i}(t) & := & Q_{i,\mu}  b_{0,i}^T y_i(t) [g\circ f_i](t) \:  [\tilde{q} \circ h](t), \quad \forall t \geq 0. 
\end{eqnarray} 
As in the proof of Proposition \ref{prop:bound:U}, we next show that  for each $i \in \llbracket 1 , \mu \rrbracket$, there exist continuous functions $c_{i,l} : \reels^{\mu -l}_{>0 } \to \reels_{>0} $ , $l \in \llbracket 1 , i \rrbracket$, such that, for all $t \geq 0$,
\begin{equation}
\label{bound_U_i_k+1:mi}
\abs{U_{1i}^{(k+1)}(t) } \leq  \sum\limits_{l=1 }^{i} a_l c_{i,l}(a_\mu , \ldots , a_{l+1}),
\end{equation}
$c_{i,\mu}$ is actually a constant independent of $a_\mu$, we write it as $c_{i , \mu} (a_{\mu},a_{\mu+1}) $ for the sake of notation homogeneity. For $i \in \llbracket 1 , \mu \rrbracket $, we apply Leibniz's rule to \rref{def:U_i_t_1:mi} and obtain that the $(k+1)$-th time derivative of $U_{1i}(\cdot)$ is given, for all $t \geq 0$, by 
\begin{eqnarray*}
U_{1i}^{(k+1)}(t) & = & a_i Q_{i+1,\mu} \left(   \sum\limits_{l_1 = 0}^{k+1 } \sum\limits_{l_2 = 0}^{l_1 } \binom{k+1}{l_1} \binom{l_1}{l_2}[\tilde{q} \circ h]^{(k+1-l_1)}(t) \,   [g\circ f_i]^{(l_2)}(t) \, b_{0,i}^T y_i^{(l_1-l_2)}(t) \right).
\end{eqnarray*} Then, to get \rref{bound_U_i_k+1:mi}, it is sufficient to show that :
\begin{itemize}
\item[a)] there exists $C >0$ such that, for any $\tilde{l} \in \llbracket 0,k+1 \rrbracket$ and for all $t \geq 0$, 
$$ \abs{[\tilde{q} \circ h]^{(\tilde{l})}(t)} \leq C [\tilde{q} \circ h](t). $$
\item[b)] for each $i \in \llbracket 1 , \mu \rrbracket$, there exist $ \Psi_i, \: \Theta_{i}, \Phi_i :  \reels^{\mu - i}_{>0} \to \reels_{>0} $, and $v_{i,j} :  \reels^{\mu - j}_{>0} \to \reels_{>0}$ for $j \in \llbracket 1 , i \rrbracket$ such that, for any $\tilde{l} \in \llbracket 0,k+1 \rrbracket$ and for all $t \geq 0$, 
$$ \norme{y_i^{(\tilde{l})}(t)  } \leq  \overline{\Psi}_i(a_\mu , \ldots , a_{i+1})  \sum\limits_{l  =i}^{\mu }  \norme{y_l(t)} + \Theta_{i}(a_\mu , \ldots , a_{i+1}) \norme{z(t)} + \overline{\Phi}_i(a_\mu , \ldots , a_{i+1}) + \sum\limits_{l  =1}^{i} a_l  \tilde{v}_{l ,i} (a_\mu , \ldots , a_{l+1}). $$
\item[c)] for each $i \in \llbracket 1 , \mu \rrbracket$, there exist $ \Gamma_i, \theta_i :  \reels^{\mu - i}_{>0} \to \reels_{>0} $, and $\Gamma_{i,j} :  \reels^{\mu - j}_{>0} \to \reels_{>0}$ for $j \in \llbracket 1 , i \rrbracket$ such that, for any $\tilde{l} \in \llbracket 0,k+1 \rrbracket$ and for all $t \geq 0$, 
$$\abs{[g\circ f_i]^{(\tilde{l})}(t)} \leq [g \circ f_i](t) \Big( \Gamma_i(a_\mu , \ldots , a_{i+1}) + \sum\limits_{l  =1}^{i} a_l  \tilde{v}_{l ,i} (a_\mu , \ldots , a_{l+1}) + \theta_i(a_\mu , \ldots , a_{i+1}) \norme{z(t)}^{2 \tilde{l}} \Big). $$
\end{itemize}

%
We now establish $a)$. From an argument of induction using differentiation of $z$-subsystem \rref{sys:mi:z:g} coupled with the fact that $\kappa$ is $p$-bounded feedback law, it can easily be shown that there exist $C_0, C_1 >0$ such that for any $\tilde{l} \in \llbracket 1, k+1 \rrbracket $ and for any $t \geq 0$,  $$\norme{z^{(\tilde{l})}(t)} \leq C_0 + C_1 \norme{z(t)}.$$ Using the Leibniz rule, it can be establish that there exist $\tilde{C}_0, \tilde{C}_1 >0$ such that, for any $\tilde{l} \in \llbracket 1, k+1 \rrbracket $, $$\abs{h^{(\tilde{l})}(t)} \leq  \tilde{C}_0 + \tilde{C}_1 \norme{z(t)}^2,$$ for all $t \geq 0$. Thanks to Fa\'a Di Bruno Formula (Lemma \ref{lem:fa_di}) applied to $[q\circ h]$, item $a)$ follows.

We now deal with item $b)$. From Lemma \ref{lem:struct:l} and an induction argument using differentiation of system \rref{sys:mi:x:g}, one can obtain the following statement: for any $l_1 \in \llbracket 1 , k+1 \rrbracket $, $i \in \llbracket 1 , \mu  \rrbracket $, there exist continuous functions $\overline{\Psi}_{l_1,i,l}  :  \reels^{\mu - i}_{>0} \rightarrow \reels_{>0}$, $l \in \llbracket i+1 , \mu  \rrbracket$ , $ \overline{\Phi}_{l_1,i,l} : \reels^{\mu - i}_{>0} \rightarrow \reels_{>0}, \quad l \in \llbracket 0 , p  \rrbracket $,  $\overline{\Theta}_{l_1,i,l} : \reels^{\mu - i}_{>0} \rightarrow \reels_{>0}, \quad l \in \llbracket 0 , p  \rrbracket$, and $\overline{\Xi}_{l_1,i,l} : \reels^{\mu - i}_{>0} \rightarrow \reels_{>0}, \quad l \in \llbracket 0 , p  \rrbracket$, such that, for all $t \geq 0$, 
\begin{eqnarray*}
\norme{y_i^{(l_1)}(t)}&  \leq &  \sum\limits_{l  =i}^{\mu } \overline{\Psi}_{l_1,i,l}(a_\mu , \ldots , a_{i+1}) \norme{y_l(t)} + \overline{\Theta}_{l_1,i,l}(a_\mu , \ldots , a_{i+1}) \norme{z(t)} \\ & + &  \sum\limits_{l  =0}^{l_1-1}  \overline{\Phi}_{l_1,i,l}(a_\mu , \ldots , a_{i+1}) \abs{U_1^{(l)}(t)} + \overline{\Xi}_{l_1,i,l} (a_\mu , \ldots , a_{i+1}) \norme{U_2^{(l_1)}(t)}.
\end{eqnarray*} So, using the inductive hypothesis and the fact that $\kappa$ is a $p$-bounded feedback law, one can obtain item $b)$.
%
%

Proceeding as in Proposition \ref{prop:bound:U}, one can get item $c)$. This ends the proof of Proposition \ref{prop:bound:U:mi}.

\section{Appendix}
\subsection{Proof of Lemma \ref{lem:SISS:int} }

Let $\epsilon >1 $ and $\beta >0$. We first prove forward completeness of 
\begin{equation}
\label{sys:int:pp}\dot{x} = - \beta \frac{x} {(1+ x^2)^{1/2}} +d_1
\end{equation}
in response to any locally bounded function $d_1(\cdot) $. For this, let $V(x):= x^2 /2$. Its derivative along trajectories of \rref{sys:int:pp} satisfies
\begin{equation}
\label{Lyap:D:int:lem}
\dot{V}(x) = - \beta  \frac{x^2}{(1+ x^2)^{1/2}}   + x^T d_1(t) .
\end{equation}
Then, a straightforward computation leads to $\dot{V}(x) \leq  V(x)  + d_1(t)^2$
and forward completeness follows using classical comparison results. Moreover when $d_1=0$, \rref{Lyap:D:int:lem} ensures that the origin of \rref{sys:int:pp} is G.A.S.

We then prove  the $ SISS_L (  \beta / 2 , \frac{ 2 \epsilon }{ \beta } )$ property of the system \rref{sys:int:pp} with respect to $d_1(\cdot)$.
Given $\delta \leq   \beta / 2 $, let $d_1$ be a bounded measurable function on $\reels_{\geq 0}$ eventually bounded by $\delta$. Since the system is forward complete, we can consider without loss of generality that
$d_1(t) \leq \delta $ for all $t \geq 0$. From \rref{Lyap:D:int:lem} and the fact that $(1+ x^2)^{1/2}  \leq 1 + \abs{x }$, one can obtain that
\begin{equation*}
\dot{V}(x) = - \beta  \frac{x^2}{(1+ x^2)^{1/2}}   + \frac{1 }{(1+ x^2)^{1/2}} ( \abs{d_1(t) } \abs{ x } +   \abs{d_1(t) } x^2 ).
\end{equation*}
Observing that 
\begin{equation}
 \frac{  \abs{d_1(t) } x^2 }{(1+ x^2)^{1/2}}  \leq   \frac{ \beta  x^2}{2(1+ x^2)^{1/2}}   ,
\end{equation}
it follows that
\begin{equation}
\dot{V}(x) \leq  -  \beta \frac{\abs{x} }{(1+ x^2)^{1/2}} \big( \abs{x} - \frac{ 2}{ \beta} \delta  \big) .
\end{equation} 
Consequently, $\dot{V} < 0$ whenever $ \abs{x} >  \frac{ 2 \delta }{ \beta}$. It follows that every trajectory of \rref{sys_int_pert_2} eventually enters and remains in the set $S= \lbrace x \in \reels \: : \: x^2 \leq \epsilon^2 ( \frac{ 2 \delta }{  \beta })^2 \rbrace$ (indeed, $\dot{V} < 0$ for all $x \notin S$ and $x \in \partial S$). Thus Lemma \ref{lem:SISS:int} can be easily established.

\subsection{Proof of Lemma \ref{Lem:osci}}

Let $\omega >0$. Given any $0 <\beta < 1$, let $A_{\beta} := \omega A_0 - \beta b_0 b_0^T $, which is Hurwitz since $A_0$ is skew-symmetric and $(A_0,b_0)$ is controllable. Therefore there exists a symmetric positive definite matrix $P_{\beta}$ satisfying the following Lyapunov equation 
\begin{equation}
\label{eq:lyap:osci}
P_{\beta} A_{\beta} + A_{\beta}^T P_{\beta} = -\identity_2.
\end{equation} A simple computation gives
\begin{equation*}
P_{\beta} = \begin{pmatrix}
\frac{\beta}{2 \omega^2}+ \frac{1}{\beta} & \frac{1}{2 \omega} \\
\frac{1}{2 \omega} & \frac{1 }{\beta}
\end{pmatrix}.
\end{equation*} 
The smallest and largest eigenvalues of $P_{\beta}$ denoted by $\underline{\sigma}_{\beta}$ and $\overline{\sigma}_{\beta}$ respectively are given by
\begin{eqnarray*}
\underline{\sigma}_{\beta} := \beta \norme{P_{\beta} b_0}^2 - \frac{\beta }{2 \omega } \norme{P_{\beta} b_0}, \\
\overline{\sigma}_{\beta}  := \beta \norme{P_{\beta} b_0}^2 + \frac{\beta }{2 \omega } \norme{P_{\beta} b_0},
\end{eqnarray*} with
\begin{equation*}
\norme{P_{\beta} b_0} = \sqrt{\frac{1}{4 \omega^2}+ \frac{1}{\beta^2}}.
\end{equation*}

Define $V_{ } : \reels^2 \to \reels_{\geq 0}$ as 
\begin{equation}
\label{Lyap_osci}
V (x):= x^T P_{\beta} x + \frac{ (\overline{\sigma}_{\beta} + \underline{\sigma}_{\beta} )  }{3} \left( (1 + \norme{x}^2 )^{3/2}  -1 \right), \quad \forall x \in \reels^2.
\end{equation} 
Given $C > 1$, let $\alpha_1$ and $\alpha_2$ be class $\kk_{\infty}$ functions given by
\begin{eqnarray*}
\alpha_{1}(r ) & :=  & \frac{(\overline{\sigma}_{\beta} + \underline{\sigma}_{\beta} ) }{C}  \max \lbrace r^2 , r^3 \rbrace , \\
\alpha_{2}(r ) & := & C (\overline{\sigma}_{\beta} + \underline{\sigma}_{\beta} )   \max \lbrace r^2 , r^3 \rbrace .
\end{eqnarray*} 

There exists $C>1$ such that
\begin{equation*}
\alpha_1(\norme{x}) \leq V(x) \leq \alpha_2(\norme{x}), \quad \forall x \in \reels^2 .
\end{equation*} Moreover, there exists a constant $M>0$, independent of $\beta$, such that 
\begin{equation}\label{assert:lyap:osci}
\alpha_1^{-1} \circ \alpha_2 (r) \leq M r, \quad  \forall r \geq 0.
\end{equation}

Proceeding as in the proof of Lemma~\ref{lem:SISS:int}, forward completeness of 
\begin{equation}\label{sys:osci:ppp}
\dot{x}= \omega A_0 x - \beta  b_0 \frac{b_0^T x}{(1+ \norme{x}^2)^{1/2}}+d_1
\end{equation} 
can easily be derived in response to any locally measurable bounded function $d_1$. 
We next show that  the system \rref{sys:osci:ppp} is $SISS_L(\beta \Gamma  , N / \beta )$ with respect to $d_1$, for some $N>0$ and with
\begin{equation}
\label{pr:elm:osci:Delta}
\Gamma := \frac{1}{8\big(  \frac{1}{4 \omega^2} +1 \big)}.
\end{equation}
Since \rref{sys:osci:ppp} is forward complete, we can assume without loss of generality that $d_1$ satisfies $\norme{d_1(t) } \leq \delta , \quad \forall t \geq 0$, for some $\delta \leq   \beta \Gamma $. Consider the Lyapunov function $V : \reels^2 \to \reels$ defined in \rref{Lyap_osci}. By noticing that \rref{sys:osci:ppp} can be rewritten as
\begin{equation*}
\dot{x}= A_{\beta} x + \beta b_0 b_0^T x \left( 1  - \frac{1 }{(1+ \norme{x}^2)^{1/2}}\right)+  d_1,
\end{equation*} 
one gets that the time derivative of $V $ along trajectories of \rref{sys:osci:ppp} satisfies 
\begin{align*}
\dot{V}= &  x^T  P_\beta \left( A_{\beta} x + \beta b_0 b_0^T x \Big( 1  - \frac{1}{(1+ \norme{x}^2)^{1/2}} \Big)+  d_1 \right) +  \left(x^T A_{\beta}^T + \beta b_0^T  b_0^T x ( 1  - \frac{1 }{(1+ \norme{x}^2)^{1/2}})+  d_1^T \right)  P_\beta x  \\ 
 &+   (\overline{\sigma}_{\beta} + \underline{\sigma}_{\beta} )(1+ \norme{x}^2)^{1/2} \left(  - \beta  \frac{ (b_0^T x)^2}{(1+ \norme{x}^2)^{1/2}}+  x^T d_1 \right).
\end{align*} Since $ P_\beta$ is a symmetric matrix satisfying the Lyapunov equation \rref{eq:lyap:osci}, it follows that
\begin{align*}
\dot{V} = & -  \norme{x}^2 + 2  \beta  x^T P_\beta b_0  b_0^T  x   \Big( 1  - \frac{1 }{(1+ \norme{x}^2)^{1/2}} \Big)+ 2 x^T  P_\beta  d_1    -  \beta   (\overline{\sigma}_{\beta} + \underline{\sigma}_{\beta} )(b_0^T x)^2+  (\overline{\sigma}_{\beta} + \underline{\sigma}_{\beta} ) (1+ \norme{x}^2)^{1/2}  x^T d_1 .
\end{align*}
By completing the squares it holds that, for all $t\geq 0$,
\begin{equation*}
\abs{ 2 \beta  x^T  P_\beta b_0  b_0^T  x  \Big( 1  - \frac{1 }{(1+ \norme{x}^2)^{1/2}} \Big)} \leq \frac{\norme{x}^2}{2} + 2  \beta^2 \norme{ P_\beta b_0}^2 (b_0^T x)^2.
\end{equation*}
Therefore, one can get that
\begin{equation*}
\dot{V} \leq  - \frac{ 1}{2} \norme{x}^2  + 2  x^T P d_1 + 2 \beta \norme{Pb_0}^2 (1+ \norme{x}^2)^{1/2}   x^T d_1 .
\end{equation*}
Using the fact that $(1+ \norme{x}^2)^{1/2} \leq 1 + \norme{x}$ for all $x \in \reels^2$, and exploiting \rref{pr:elm:osci:Delta}, it follows that
\begin{equation*}
\dot{V} \leq  - \frac{ 1}{4} \norme{x}^2  +  2 \norme{x} \delta \left( 2 \beta \norme{P_{\beta} b_0}^2 + \frac{\beta }{2 \omega } \norme{P_{\beta} b_0} \right).
\end{equation*} 
Consequently, it holds that $\dot{V}  < 0 $ whenever $\norme{x} >  8 \delta  \big( 2\beta \norme{P_{\beta} b_0}^2 + \frac{\beta }{2 \omega } \norme{P_{\beta} b_0} \big)   $. Let $\mu > 1$ and set $r := 8 \mu  ( 2 \beta \norme{P_{\beta} b_0}^2 + \frac{\beta }{2 \omega } \norme{P_{\beta} b_0})  $. Define $S:=\lbrace x \in \reels^2 \: : \:  V(x) \leq \alpha_2 (r \delta ) \rbrace$.  If $x \notin S$ then $\norme{x} > r \delta $. Consequently, any trajectory eventually enters and stay in $S$. Moreover, we have that $\alpha_1( \norme{x(t)} ) \leq_{ev} V(x(t)) \leq \alpha_2(r \delta )$ . From \rref{assert:lyap:osci}, it follows that $\norme{x(t)} \leq_{ev} rM \delta $. Moreover, one can see that there exists a constant $D>0 $ such that for any $\beta \leq 1$ we have $r \leq \frac{D}{\beta}$. So we obtain 
\begin{equation*}
 \norme{x(t)} \leq_{ev} \frac{N \delta}{\beta}, 
\end{equation*} for some $N>0$, which concludes the proof.

\subsection{Fa\`a Di Bruno's Formula}

\begin{lem}[Fa\`a Di Bruno's formula, \cite{fdb}, p. 96]
\label{lem:fa_di}
For  $k\in\mathbb N$, let $\phi\in C^{k}( \reels_{\geq 0} , \reels )$ and $\rho\in C^{k}( \reels , \reels )$. Then the $k$-th order derivative of the composite function $\rho \circ \phi$ is given by
\begin{equation*}
[ \rho \circ \phi ]^{(k)}(t) =  \sum\limits_{a=1}^k \rho^{(a)} (\phi(t)) B_{k,a}\Big(\phi^{(1)}(t), \ldots , \phi^{(k-a+1)}(t)\Big),
\end{equation*}
where $B_{k,a}$ is the Bell polynomial given by
\begin{align*}
B_{k,a}\Big(\phi^{(1)}(t), \ldots ,  \phi^{(k-a+1)}(t)\Big)\hspace{-1mm}:=\hspace{-2mm}\sum\limits_{\delta \in \pcal_{k,a}} \hspace{-1mm}c_{\delta} \hspace{-1mm}\prod\limits_{l=1}^{k-a+1} \left( \phi^{(l)}(t) \right)^{\delta_l},
\end{align*}
where $\pcal_{k,a}$ denotes the set of $(k-a+1)-$tuples $\delta :=(\delta_1 , \delta_2, \ldots , \delta_{k-a+1})$  of positive integers satisfying 
\begin{align*}
&\delta_1 + \delta_2 + \ldots +\delta_{k-a+1} = a,\\
&\delta_1 +2 \delta_2 + \ldots + (k-a+1) \delta_{k-a+1} = k,\\
&c_{\delta}:=\frac{k!}{\left(\delta_1 ! \cdots \delta_{k-a+1}! (1!)^{\delta_1} \cdots ((k-a+1)!)^{\delta_{k-a+1}}\right)}.
\end{align*}
\end{lem}

\bibliographystyle{IEEEtran}
\bibliography{IEEEabrv,biblio}
\end{document}